\documentclass[useAMS,usenatbib]{mn2e}

\usepackage{times}

\usepackage{graphicx}
\usepackage{times}
\newcommand{\cthead}[1]{\multicolumn{1}{c}{#1}}
\newcommand{\kss}{km~s$^{-1}$ }
\newcommand{\ks}{km~s$^{-1}$}

\title{New 9.9-GHz methanol masers}
\author[M. A. Voronkov et al.]{M. A. Voronkov$^{1,2}$\thanks{E-mail:
Maxim.Voronkov@csiro.au}, J.L. Caswell$^1$, S.P. Ellingsen$^{3}$, A.M. Sobolev$^4$\\
$^{1}$Australia Telescope National Facility, CSIRO Astronomy and Space Science, PO Box 76, 
Epping, NSW 1710, Australia\\
$^{2}$Astro Space Centre, Profsouznaya st. 84/32, 117997 Moscow, Russia\\
$^{3}$School of Mathematics and Physics, University of Tasmania, GPO Box
252-37, Hobart, Tasmania 7000, Australia\\
$^{4}$Ural State University, Lenin ave. 51, 620083 Ekaterinburg, Russia\\}
\begin{document}

\date{}

\pagerange{\pageref{firstpage}--\pageref{lastpage}} \pubyear{2009}

\maketitle

\label{firstpage}

\begin{abstract}
The Australia Telescope Compact Array (ATCA) has been used 
to make the first extensive search for the class~I methanol masers
at 9.9~GHz. In total, 48 regions of high-mass star formation were observed.
In addition to masers in W33-Met (G12.80$-$0.19) and G343.12$-$0.06
(IRAS 16547$-$4247) which have already been reported in the literature, 
two new 9.9-GHz
masers have been found towards G331.13$-$0.24 and  G19.61$-$0.23.
We have determined absolute positions (accurate to roughly a second of arc) for all 
the detected masers and suggest that some class~I masers may be 
associated with shocks driven into molecular clouds by
expanding H{\sc ii} regions.  Our observations also imply that the evolutionary stage of 
a high-mass star forming region when the class~I masers are present
can outlast the stage when the class~II masers at 6.7-GHz are detectable,
and overlaps significantly with the stage when OH masers are active.
\end{abstract}


\section{Introduction}

Methanol masers are well established probes of high-mass star formation. 
They are divided into two categories
first defined by \citet{bat87}: class~II masers (of which the 6.7~GHz is the best
known and usually strongest) are associated with millimetre
and infrared sources \citep[e.g.,][]{hil05} and reside in the close environment of 
high-mass young stellar objects (YSOs).  Class~I masers (of which the 44 and 95~GHz
are commonly observed) are often found apart from the strong continuum 
and infrared sources and can be separated by up to a parsec from the 
YSO responsible for excitation \citep[e.g.,][]{kur04, vor06, cyg09}. 

Theoretical calculations are able to explain this empirical classification and strongly
suggest that the pumping process of class~I masers is dominated by collisions 
with the molecular hydrogen \citep[see, e.g.,][]{lee73,sob83,vor99}
in contrast to class~II masers, which are pumped by radiative excitation 
\citep[see, e.g.,][]{wil85, sob94, sut01}. The two pumping mechanisms were shown to be 
competitive \citep[see, e.g.,][]{men91,cra92,vor99,vor05b}. For example, strong radiation 
from a nearby infrared source quenches class~I masers and increases the strength of 
class~II masers \citep{vor05b}.  Although it has been demonstrated by theoretical calculations that a weak class~II maser at 6.7~GHz can coexist with a bright class~I emission under special conditions \citep{vor05b},
bright masers of different classes residing in the same volume of gas are widely accepted as mutually
exclusive.  However, on larger scales, they are often observed to coexist in the same star forming region within less than a parsec of each other.

Class~I methanol masers are relatively poorly studied. The common consensus is that the 
majority of class~I masers trace interface regions between outflows and molecular gas,
although direct observational evidence of this has been obtained for a limited number of 
sources only \citep[e.g.,][]{pla90,kur04,vor06}. An alternative scenario, which involves 
cloud-cloud collisions, may be realised in some sources \citep{sob92,meh96,sal02}.
The common ingredient of these two scenarios is the presence of shocks. There is 
observational evidence that the methanol abundance is significantly increased in the shock
processed regions \citep[e.g.][]{gib98,sut04}. The gas in such regions is heated and compressed
increasing the frequency of collisions with molecular hydrogen and, therefore,
providing a more efficient pumping \citep[e.g.][]{sut04}. 
\citet{che09} demonstrated statistically the association of 
class~I masers with the shocks traced by extended features showing a prominent excess of 
the 4.5-$\mu$m emission in the images obtained
with the Spitzer Space Telescope's Infrared Array Camera (IRAC) also known as
extended green objects \citep[EGOs;][]{cyg08}. 

To date, there is observational evidence that more than 20 methanol transitions can produce
class~I masers, more than half of which belong to the J$_2-$J$_1$~E series in the 
25$-$30~GHz frequency range \citep{vor99,mul04}. The observational properties of
different class~I masers are not the same. Some masers, e.g. at 44 and 95~GHz, are known 
to be quite ubiquitous with about 100 sources known to date 
\citep[see, e.g.,][]{has90,sly94,val00,kur04,cyg09}. On the other hand, there are other transitions
with very few known maser sources.  
An example of such a transition showing rare maser activity is the $9_{-1}-8_{-2}$~E methanol 
transition at 9.9~GHz. The rarity of these masers is likely due to  the
strong dependence  of the maser brightness on the physical conditions,
especially in the requirement of  higher typical temperatures and densities for these 
masers to form \citep{sob05}.  

\citet{sly93} conducted the only search for the 9.9-GHz masers reported in the literature so far.  
However, they observed just 11 targets which were largely well known regions of high-mass
star-formation where methanol masers of either class had previously been reported and  
discovered a single 9.9-GHz maser (W33-Met  also known as G12.80$-$0.19).  
In this paper we report the results of the search for the 9.9-GHz masers 
towards 48 independent positions. The majority of targets, 46 sources in total, were known 
class~I masers at 44 and/or 95~GHz drawn from \citet{sly94}, \citet{val00}, \citet{kur04}, and 
\citet{ell05}. The remaining two sources were included in order to cover all known
star forming regions located south of declination of $-$20$^\circ$, which have a periodically 
variable class~II methanol maser at 6.7-GHz according to \citet{goe04}. Little is known about 
the physics of periodic variability of such masers. Therefore, a detection of the 9.9-GHz maser
(which is very sensitive to the physical conditions and has a different pumping to the 6.7-GHz maser)
in the same source might shed light on this enigmatic phenomenon.

\section{Observations}
\label{obs_section}

\begin{table*}
\caption{Dates of observations, a summary of array configurations and decorrelation estimates. More details on the ATCA
configurations are available on the web ({\it http://www.narrabri.atnf.csiro.au/observing/configs.html}). Note, that
antenna CA01 did not perform well at this frequency in 2005/2006. It is involved, in particular, in the
shortest spacing of the 6C array configuration. Therefore, the shortest baseline was not as sensitive
as the other baselines in March observations. Antenna CA06 did not observe in 2008 sessions.}
\label{obstab}
\begin{tabular}{@{}rcllllrrr}
\cthead{UT  Date} & \cthead{Array} & \multicolumn{3}{c}{Baseline length}  & \multicolumn{4}{c}{Decorrelation estimate} \\
\cthead{of} & \cthead{configuration}&\cthead{min} & \cthead{max} & \cthead{max} & \cthead{Source} &
\multicolumn{2}{c}{Separation} & \cthead{Estimated}\\
\cthead{observations}  &  &  & \cthead{no CA06} & \cthead{with CA06}  & \cthead{and} &\cthead{angular} &
\cthead {temporal} & \cthead{factor}\\
& & \cthead{(m)}&\cthead{(m)}&\cthead{(m)}  &\cthead{calibrator} & \cthead{(deg)} & \cthead{(min)} & \\
\hline
15 Dec 2005 & 6A & 337 & 2923 & 5939  & 1059-63\hphantom{1} (odd+even) & 0\hphantom{.0} & 23.8 (2) & 3.8\hphantom{18} (7) \\
\multicolumn{5}{c}{} & 1613-586 (odd+even) & 0\hphantom{.0} & 21\hphantom{.0} (5)& 5.7\hphantom{18}  (5)  \\
16 Dec 2005 &   &   &  &  &  1613-586 (odd+even) & 0\hphantom{.0} & 18\hphantom{.0} (3) &  6.8\hphantom{18} (4) \\
17 Dec 2005 &  &  &  &  & 1613-586 (odd+even) & 0\hphantom{.0} & 17\hphantom{.0} (3) & 1.88\hphantom{1}  (3)  \\
25 Mar 2006 & 6C & 153 & 2786 & 6000 & 1646-50\hphantom{1} (odd+even) & 0\hphantom{.0} & 27.8 (1)& 1.179 (4) \\
\multicolumn{5}{c}{} & 1710-269 + 1730-130 & 14.6 & 20.1 (7) & 1.25\hphantom{1} (1) \\
26 Mar 2006 &   &   &   &  & 1646-50\hphantom{1} (odd+even) & 0\hphantom{.0} & 20.4 (2) & 1.14\hphantom{1} (2) \\
\multicolumn{5}{c}{}& 1710-269 (odd+even) & 0\hphantom{.0} & 21\hphantom{.0} (1) & 1.250 (9) \\
2 Jun 2008 & EW352 & 31 & 352 & none & 1646-50 (odd+even) & 0\hphantom{.0} & 17\hphantom{.0} (2) & 1.004 (5) \\
20 Jul 2008 & H214 & 92 & 247 & none & 1646-50 (odd+even) & 0\hphantom{.0} & 9\hphantom{.0}
(2) & 1.036 (2)\\
\multicolumn{5}{c}{} & 1613-586 (odd+even) & 0\hphantom{.0} & 9\hphantom{.0}
(2) &  1.065 (3)\\
\multicolumn{5}{c}{} & 1646-50 + 1613-586 & 9.3 & 1.8 (1) & 1.16\hphantom{1} (2)\\
\hline
\end{tabular}
\end{table*}
Observations were made with  the Australia Telescope Compact Array (ATCA) in 7 time allocations scheduled
from December 2005 to July  2008 (the archive project code is C1466). The details of these observing sessions 
along with the array configurations used and the range of baseline lengths are summarised in 
the first five columns of Table~\ref{obstab}. Each source was observed in several (typically 6) cuts
of 10 minutes each to get a good hour angle coverage. The data
reduction was performed using  the {\sc miriad} package (12 June 2007 release) following standard
procedures (with the exception of bandpass calibration). The bandpass calibration was achieved by fitting a low-order 
polynomial to the spectra of the bandpass calibrator (1921-293,  except for the observations
on December 17 and June 2, when 0537-441 and 1253-055 were observed, respectively) 
using the {\sc uvlin} and {\sc mfcal} tasks.
This approach for bandpass calibration is usually superior and adds less noise than an independent
solution for each 
spectral channel, if one tries to detect a weak narrow line in the presence of a strong continuum,
which is the case for many sources in our sample. 

The regular observations were preceded by a short
test project (the archive project code is CX075) using Director's time in December 2004
and January 2005. It aimed at testing the viability of observations at 9.9~GHz, which is outside the 
nominal range 
of the ATCA X-band receivers (8.0 to 9.2~GHz), but yielded the serendipitous discovery of a new 
9.9~GHz maser in G343.12-0.06 (IRAS 16547-4247). This source was then investigated in 
detail in a separate paper \citep{vor06}. The receiver performance was found to vary significantly 
from antenna to antenna at 9.9~GHz due to the proximity of the band edge. In particular,
baselines including antenna CA01 had  approximately 3 times higher noise than other baselines in
most sessions. As the masers are unresolved sources for the ATCA, this inhomogeneity in the
sensitivity of individual baselines is taken care of automatically by the imaging process.

We adopted an astronomical measurement of the rest frequency for the $9_{-1}-8_{-2}$~E
methanol transition, 9936.201$\pm$0.001~MHz \citep{vor06}. The uncertainty corresponds to  
0.03~\kss of the radial velocity. The adopted rest frequency lies within the uncertainty of the 
laboratory measurement of \citet{mul04} equivalent to 0.12~\ks. The correlator was
configured with 1024 spectral channels across a 4~MHz bandwidth, providing a spectral resolution of
0.12~\ks. In addition, the spare correlator capacity was used to observe the 8.6 GHz continuum emission 
(with the 128-MHz bandwidth) simultaneously with the line measurement. 
Although two orthogonal linear polarisations
have been recorded, the correlator configuration
used in the project did not allow us to calibrate the instrumental polarisation and hence to 
determine polarisation properties of the observed emission. During the data reduction,
the orthogonal polarisations were converted to Stokes-I. Therefore, all flux densities  presented in 
the paper  are  not affected by polarisation to the first order.

All image cubes were constructed from the visibility data after continuum subtraction. The natural
weights were used providing a typical synthesised beam 
size of around a few seconds of arc.  Due to software limitations the search area was split into 9
sub-cubes with a 50\% overlap. Each sub-cube containing 512$\times$512 spatial pixels of 
0.5$\times$0.5 arcsec and 1024 spectral channels was then processed separately. This strategy covered
a rectangular area of 8.3~arcmin in size. The half power width of the primary beam is
5.1~arcmin at this frequency. An automated spectral line finder (Voronkov, in prep.) based on the 
ATNF Spectral line Analysis Package ({\it http://www.atnf.csiro.au/computing/software/asap}) was
used to search for sources in the dirty spectral cubes.\footnote{Strictly speaking, sidelobes are only
caused by real sources. Therefore, no deconvolution is required unless there is a detection.}
The spectral line finder applied a simple statistical criterion of at 
least 3 consecutive spectral channels above the threshold of 3$\sigma$ (with respect to the
noise in the local vicinity of the tested channels) to claim a detection. The 3-dimensional search routine
formed slices along the spectral axis for each spatial pixel, ran the 1-dimensional algorithm and
combined the results into sources. The sources which had less than 4 spatial pixels with
detected emission (half power area of the main lobe of synthesised beam) were discarded.   
Due to statistical noise fluctuations, the algorithm 
produced of the  order of 10 spurious sources per cube, each of which was examined manually.
The same source detection software was used and thoroughly tested in the Parkes 6-GHz 
Multibeam survey  \citep[hereafter MMB survey;][]{gre09}. 
In addition to automated software, both single baseline and
averaged uv-spectra (visibility amplitude versus frequency) were inspected 
by eye to search for possible broad thermal components, which may not be imaged adequately 
with the uv-coverage provided by the sparse ATCA configurations. Due to the low dynamic range in the dirty image cubes the automated software reported a number of detections around the field of 
view for  each real maser.  In this case, a deconvolved cube was made using the {\sc clean} task and the 
source parameters were measured using the {\sc imfit} task of {\sc miriad}.  
A self-calibration was performed and a new cube was constructed after the absolute position had been determined. 
The spectra were extracted from such self-calibrated cubes using the {\sc imspec} task. 
An additional search for weaker masers was performed in the parts of the self-calibrated cube 
unaffected by the main maser feature using the same approach as for the original
cubes. However, we did not find any additional weak maser features.

During
the first two observing allocations of the regular observations in December 2005 the atmospheric conditions
were very unstable with thunderstorms
present in the area causing an interruption once on December 15. Observations through 
adverse weather, aggravated by the sparse array configuration, resulted in an increased level of
decorrelation due to a poor phase stability. As this decorrelation directly affects
the flux density limit for all non-detections, we have attempted to estimate it using the calibrator data.
We used two different methods, depending on the data available for a particular observing session. On March 25 
two separate continuum calibrators 1710-269 and 1730-130 were observed using the same
frequency setup, which was chosen to achieve an appropriate velocity range for the target sources. The first source of this pair acted as an ordinary secondary calibrator allowing us
to form a calibration solution. The second source was then imaged using this calibration. Another image of the second
source was constructed following a self-calibration.  Then, the decorrelation factor was
measured as the ratio of the two flux densities obtained from each image using the {\sc imfit} task.
The second method, suitable for observations with only one calibrator, is to split the whole sequence of calibrator scans into odd and even
scans and then treat the odd ones as a calibrator and the even scans as a source. The decorrelation
factor was then obtained from the images the same manner as for the first method. 

\begin{table*}
\caption{Observation details for sources detected at 9.9 GHz. All uncertainties are given in the brackets
and expressed in the units of the least significant figure. The uncertainty of the decorrelation factor
was calculated from the uncertainties of the flux measurement.}
\label{pos_obsdetails}
\begin{tabular}{@{}lr@{}rcclcrr}
\cthead{Source} & \cthead{Date of} & \multicolumn{2}{c}{Synthesised beam} & \cthead{$1\sigma$ Noise}
&\cthead{Inspected}&\cthead{Measured} &
\multicolumn{2}{c}{Separation from calibrator}\\
\cthead{name} & \cthead{observation} & \cthead{FWHM} &\cthead{p.a.} &\cthead{level} &
\cthead{velocity range}&\cthead{decorrelation} & \cthead{angular} & \cthead{temporal} \\
& & \cthead{(arcsec)}  & \cthead{(deg)} &\cthead{(mJy beam$^{-1}$)}
&\cthead{(\ks)}& \cthead{factor} &\cthead{(deg)} & \cthead{(min)}\\
\hline
G331.13$-$0.24 & 16 Dec 2005 & 2.1$\times$1.5 & 15 & 39 & $-$124, $-$27 & 2.9\hphantom{81}(3)
& 7.0 & 5.3\hphantom{1}(1)\\
& 2 Jun 2008 & \hphantom{.}18$\times$12\hphantom{.} & \hphantom{1}3 & 35 & $-$131, $-$34 & 1.04\hphantom{1}(2) & 6.2 & 4.0\hphantom{1}(1) \\
& 20 Jul 2008 & \hphantom{.}17$\times$14\hphantom{.} & 53 & 68 & $-$150, $-$53 & 1.09\hphantom{1}(3) & 6.2 & 4\hphantom{.41}(1)  \\
W33-Met      & 25 Mar 2006 & 2.6$\times$1.0 & 14 & 44 & $-$29\hphantom{4}, $+$53 & 1.18\hphantom{1}(4) 
& 11.0 & 5.4\hphantom{1}(1) \\
G19.61$-$0.23   & 25 Mar 2006 & 3.7$\times$1.0 & \hphantom{1}7 & 43  & $-$28\hphantom{4}, $+$68  & 1.60\hphantom{1}(6)
& 13.4 & 21\hphantom{.41}(3) \\
\hline
\end{tabular}
\end{table*}

The right-hand four columns of Table~\ref{obstab} 
summarise  the results for each observing sessions, including the method and sources used, angular and
mean temporal separation between the would be target and calibrator, 
and the estimated decorrelation factor. The uncertainties are expressed in the units of the least 
significant figure. Table~\ref{obstab} shows that the decorrelation was up to 7 times worse during the 
first two observing sessions. It is worth noting that these estimates are likely to be larger than the
actual decorrelation affecting the target sources because we effectively double the length of
the calibrator cycle when splitting calibrator scans into odd and even scans.  For some observing
sessions more than one decorrelation estimate is available (Table~\ref{obstab}). The estimates are in a good
agreement  for the session on December 15, when the phase stability was bad, but are more
than 3$\sigma$ apart for the March sessions, which had a good phase stability. This is likely to be 
due to systematic effects such as different declinations  for the calibrators involved in the
estimates (which makes the average elevations different), and an additional angular separation when a pair of sources was
used instead of the odd+even method.  We used the largest decorrelation estimate obtained for
each observing session to scale the flux density limits for non-detections.

The last two sessions in 2008 (Table~\ref{obstab}) were completely devoted to 
the  9.9~GHz maser discovered towards G331.13$-$0.24 during the session on December 16 and no
new targets were observed. The goal of these repeated observation was to provide an
independent measurement of the absolute position of the maser. We found that the morphology of 
G331.13$-$0.24 was such that an independent estimate of a systematic uncertainty was 
required for an unambiguous interpretation. A typical ($1\sigma$) astrometric accuracy of ATCA (largely determined by systematic effects for masers) is usually adopted to be around 0\farcs5. 
However, the bad weather of the December 16 session is likely to have further degraded
the accuracy. Although two 2008 sessions used relatively 
compact array configurations and the synthesised beam was an order of magnitude larger, 
by observing in good weather and by using a different calibrator (1646-50 instead of 1613-586)
for which a more accurate position is available,  we have achieved a comparable
formal uncertainty of the position measurement in all three sessions.  We found that the absolute 
positions obtained in all three sessions are consistent, and we estimate the final accuracy (sigma
for the mean) to be 1 arcsecond.


\section{Results}

In addition to the new 9.9-GHz maser found in G343.12$-$0.06 during the pilot survey 
as a chance discovery \citep[results of the follow-up are reported in][]{vor06}, the regular survey
yielded three other detections, listed in Table~\ref{pos_obsdetails}. These include the known
9.9 GHz maser in W33-Met (G12.80$-$0.19) discovered by \citet{sly93}, and two new detections. 
The first six columns of the table are self-explanatory and include the date of
observation, the synthesised beam size and its position angle, the noise level in the image 
representing the detection threshold, and the velocity range searched for maser emission. The 
last three columns  are related to the decorrelation estimates.  For sources with a maser detection,
self-calibration effectively provides a measurement of the decorrelation factor. The factor
given in Table~\ref{pos_obsdetails} was calculated as the ratio of the flux density obtained
using a self-calibrated data set of the maser to that prior to the self-calibration. The values are 
below the estimates obtained for a particular observing session using the calibrator 
data (see Table~\ref{obstab}) as they are expected to be, with the exception of 
the last source, G19.61$-$0.23. For this source, the measured decorrelation is slightly higher 
than the estimate deduced from the calibrator data. This fact can probably be explained by a 
more northern location of the source than either of the calibrators used for decorrelation estimates 
and, as a result, a lower elevation on average throughout the observing run. Systematic
effects such as fit errors due to the elongated synthesised beam along the declination axis,
which is produced by the East-West arrays for the sources located near the equator, could also 
contribute to this minor discrepancy. For the purpose of comparison, the last two columns
of Table~\ref{pos_obsdetails} give the angular and temporal separation of the maser 
source from the appropriate secondary calibrator. The uncertainties are given in round 
brackets and expressed in the units of the least significant figure.

\begin{table*}
\caption{The only four known masers at 9.9~GHz. In addition to 3 sources reported in this paper
we have included G343.12$-$0.06 \protect\citep{vor06} for completeness.
The uncertainties in the absolute positions given in the table are based on the formal uncertainty of 
the fit and the uncertainty of calibrator position. The commonly adopted systematic 
uncertainty (1$\sigma$) of the absolute positions measured with ATCA is 
0\farcs5. All uncertainties in the table are given in the units of the least 
significant figure. The flux density uncertainty includes both the image noise (see Table~\protect\ref{pos_obsdetails}) 
and the accuracy of the absolute flux scale calibration. The uncertainty of the peak velocity 
is the spectral resolution.
}
\label{positive}
\begin{tabular}{@{}lrlllllr}
\cthead{Source}& \cthead{Date of}& && \multicolumn{2}{c}{Determined position} &\cthead{Peak} &\cthead{Flux} \\
\cthead{name} & \cthead{observation} &\cthead{l} & \cthead{b} &\cthead{$\alpha_{2000}$}&\cthead{$\delta_{2000}$}&
\cthead{velocity} & \cthead{density} \\
&&\cthead{(\degr)} & \cthead{(\degr)} &\cthead{($^h$~$^m$~$^s$)}&
\cthead{(\degr~\arcmin~\arcsec)}&\cthead{(\ks)} & \cthead{(Jy)}\\
\hline
G331.13$-$0.24  & 16 Dec 2005 & 331.131 & $-$0.245 & 16:10:59.76\hphantom{7} (5) &
$-$51:50:26.9\hphantom{7} (4) & $-$91.16 & 1.92 (7) \\
& 2 Jun 2008 &&& 16:10:59.51\hphantom{7} (3) & $-$51:50:26.1\hphantom{7} (3) & $-$91.15 & 3.1\hphantom{8} (1) \\
& 20 Jul 2008 &&& 16:10:59.63\hphantom{7} (3) & $-$51:50:27.3\hphantom{7} (3) & $-$91.16 &
2.6\hphantom{1} (1) \\
G343.12$-$0.06 & 16 Jun 2005 & 343.121 & $-$0.064 & 16:58:16.460 (2) & $-$42:52:25.73 (3) & $-$31.56 & 9.5\hphantom{0} (3) \\
W33-Met       & 25 Mar 2006 & \hphantom{3}12.796 & $-$0.192 & 18:14:10.897 (3) & $-$17:55:58.42 (8) & $+$32.72 & 4.3\hphantom{1} (1) \\
G19.61$-$0.23    & 25 Mar 2006 & \hphantom{3}19.608 & $-$0.232 &18:27:37.475 (2) & $-$11:56:37.77 (8) & $+$41.24 & 3.3\hphantom{1} (1) \\
\hline
\end{tabular}
\end{table*}

\begin{figure}
\includegraphics[width=\linewidth]{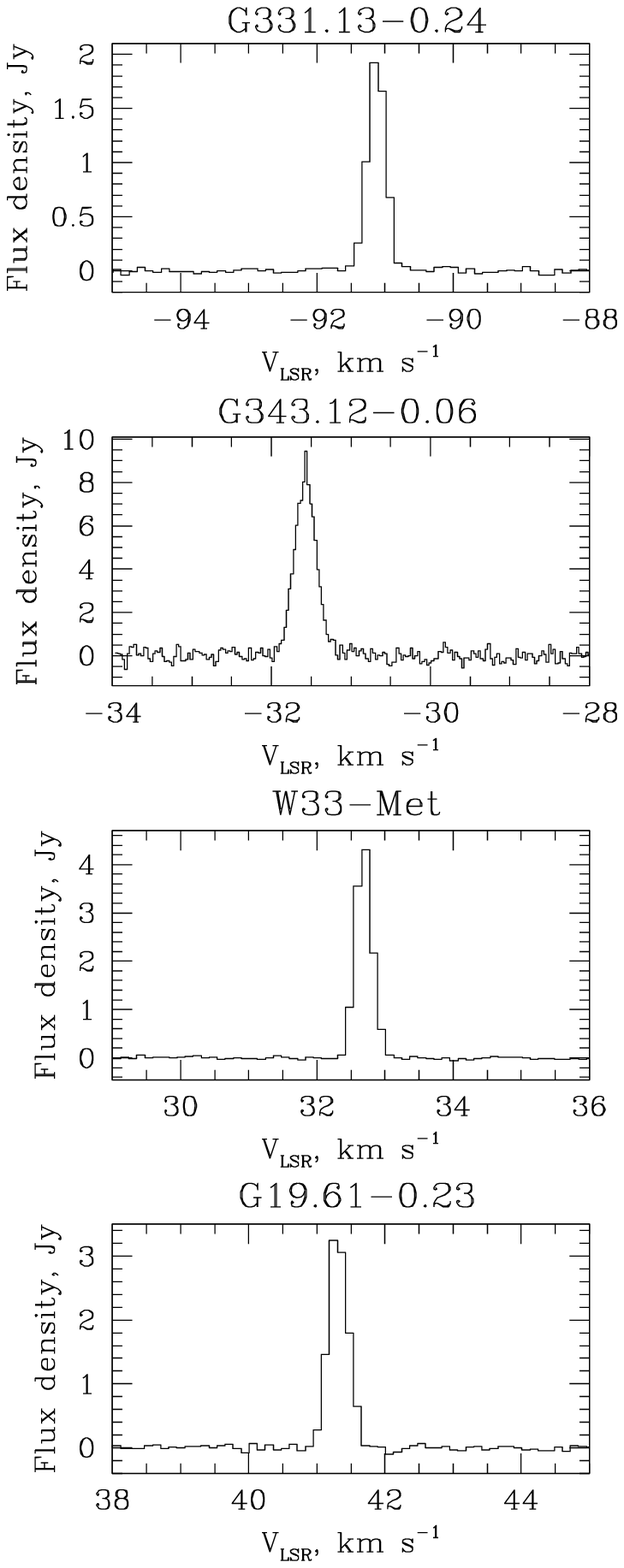}
\caption{Spectra of the masers detected at 9.9~GHz. The uncertainty in radial velocity due to
the rest frequency uncertainty is less than the spectral resolution. For completeness, we show
the spectrum of G343.12$-$0.06 \protect\citep[from][]{vor06}. Note, the spectrum of this 
source was obtained with four
times better spectral resolution, which is comparable with the rest frequency uncertainty. }
\label{spectra}
\end{figure}

The absolute position, peak velocity and flux density are summarised in Table~\ref{positive}
for all detected masers. We have added G343.12$-$0.06 \citep{vor06} so that Table~\ref{positive}
contains a complete list of all four 9.9~GHz masers known to date.
The uncertainty of the peak velocity is the spectral resolution given in
the previous section (the rest frequency uncertainty is much smaller than the spectral resolution).
The flux density uncertainties  given in Table~\ref{positive} are cumulative values including both
the image noise given in Table~\ref{pos_obsdetails} and the accuracy of the flux scale calibration,
which is typically of the order of 3\% at cm-wavelengths. The spectra of the 9.9-GHz detections are
shown in Fig.~\ref{spectra}. All masers
have simple spectra in this transition containing just one component. In contrast, most  sources
have multiple components in the spectra of the most widespread class~I
methanol maser transition at 44~GHz \citep{sly94,kog98}. This situation is similar to
that observed in G343.12$-$0.06, where the 9.9~GHz maser and other rare
class~I masers at 25 and 104~GHz are confined to one spot only, while the more widespread
95 and 84-GHz masers have been observed in a number of spots \citep{vor06}. Such a behavior
confirms the hypothesis that the 9.9~GHz masers are produced under a more restricted range of physical conditions than the widespread masers.

Figure~\ref{maps} shows positions of detected 9.9-GHz masers with respect to the 8.6-GHz continuum 
observed simultaneously. The positions
of the 6.7-GHz masers (see notes on individual sources for references) are also shown. 
The grayscale represents the Spitzer IRAC 4.5-$\mu$m images.
For one  source, G331.13$-$0.24, we used the contiuum data of \citet{phi98}. 
Both the morphology of the H{\sc ii} region and the flux density obtained in our measurement are in a good agreement with \citet{phi98}. But the map of \citet{phi98}, with better 
positional accuracy and uv-coverage than in our measurement,
reflects better the relative offset between the 6.7-GHz maser and the H{\sc ii} region. 

The list of sources without a 9.9-GHz detection is given in Table~\ref{negative}. The first
7 columns are self-explanatory and include the source name, the date of observations,
observed position and velocity range searched for maser emission as well as the synthesised
beam size (given as the full width at half maximum) and its position angle. The eighth column
contains the noise level (1$\sigma$) in the dirty cubes searched for masers. This raw value
does not take into account decorrelation due to weather, which we estimated using
the calibrator data (Table~\ref{obstab}). As mentioned above, we adopted the largest estimate
of the decorrelation factor obtained for the particular observing slot. The cube noise level scaled with
this factor is given in the ninth column of Table~\ref{negative} as the estimate of the true
1$\sigma$ detection limit.
As this limit is likely to be overestimated a bit for the majority of sources observed in a
bad weather we give the mean temporal and angular separation from the appropriate secondary calibrator in
the last columns of Table~\ref{negative}. The smaller the separation,  the lower
the actual detection limit is expected to be. The uncertainties are given in the round brackets
and expressed in the units of the least significant figure.

\begin{table*}
\caption{Sources without a 9.9-GHz detection. Flux density limit is $1\sigma$. G329.031-0.19 and G329.029-0.20 represent two close
positions within the same source.}
\label{negative}
\begin{tabular}{@{}lrlll@{}c@{}c@{}c@{}c@{}r@{}r}
\cthead{Source}&\cthead{Date of}& \multicolumn{2}{c}{Observed position} & \cthead{Velocity} & 
\multicolumn{2}{c}{Synthesised beam} & \cthead{$1\sigma$ image} &\cthead{$1\sigma$ flux}& 
\multicolumn{2}{c}{Separation from calibrator}\\
\cthead{name}&\cthead{observations}&\cthead{$\alpha_{2000}$}&\cthead{$\delta_{2000}$}&\cthead{range} & 
\cthead{FWHM} & \cthead{pa}  &\cthead{noise level} & \cthead{limit}& \cthead{angular} & \cthead{temporal} \\
&&\cthead{($^h$~$^m$~$^s$)}&\cthead{(\degr~\arcmin~\arcsec)}&\cthead{(\ks)}& \cthead{(arcsec)} &
\cthead{(deg)} & \cthead{(mJy beam$^{-1}$)}& \cthead{(mJy)} &\cthead{(deg)} & \cthead{(min)}\\
\hline
G301.14$-$0.23  & 15 Dec 2005 & 12:35:34 & $-$63:02:38 & $-$97, $-$1 & 1.1$\times$0.8 & \hphantom{$-$}39 & 76 & 430  & 10.5 & 6.6\hphantom{1}(1)\\
G305.21$+$0.21  & 15 Dec 2005 & 13:11:09 & $-$62:34:41 & $-$97, $-$1 & 1.1$\times$0.8 &\hphantom{$-$}33  & 82 & 470 & 14.5 & 6.9\hphantom{1}(2)\\
G305.25$+$0.25  & 15 Dec 2005 & 13:11:33 & $-$62:32:03 & $-$97, $-$1 & 1.3$\times$0.7 & \hphantom{$-$}34 & 76 & 430 & 14.6 & 6.8\hphantom{1}(2)\\
G305.36$+$0.20  & 15 Dec 2005 & 13:12:34 & $-$62:33:25 & $-$97, $-$1 & 1.3$\times$0.7 & \hphantom{$-$}35  & 81 & 460 & 14.7 & 17.2\hphantom{1}(1) \\
G326.475$+$0.70 & 15 Dec 2005 & 15:43:17 & $-$54:07:13 & $-$94, $+$2 & 1.3$\times$0.7 & \hphantom{$-$}24 & 76 & 430 & 6.6 & 6.5\hphantom{1}(1)\\
G326.641$+$0.61 & 15 Dec 2005 & 15:44:33 & $-$54:05:29 & $-$94, $+$2 & 1.3$\times$0.8  & \hphantom{$-$}26 & 69 & 390 & 6.5 & 6.4\hphantom{1}(1)\\
G327.392$+$0.19 & 16 Dec 2005 & 15:50:18 & $-$53:57:06 & $-$124, $-$28 & 1.6$\times$0.9 & \hphantom{$-$}20 & 68 & 460 & 6.1 & 6\hphantom{.61}(1) \\
G326.859$-$0.67 & 16 Dec 2005 & 15:51:14 & $-$54:58:05 & $-$125, $-$29 & 1.6$\times$1.0 & \hphantom{$-$0}9& 89 & 610 & 5.2 & 5.4\hphantom{1}(2)\\
G327.618$-$0.11 & 16 Dec 2005 & 15:52:50 & $-$54:03:01 & $-$124, $-$28 & 1.5$\times$1.0 & \hphantom{$-$}14 & 64 & 440 & 5.8 & 5.2\hphantom{1}(1)\\
G327.29$-$0.58  & 15 Dec 2005 & 15:53:06 & $-$54:37:06 & $-$95, $+$1 & 1.3$\times$0.7 & \hphantom{$-$}25 & 62 & 350 & 5.3 & 6\hphantom{.41}(1)\\
G328.81$+$0.64  & 15 Dec 2005 & 15:55:48 & $-$52:42:48 & $-$94, $+$2 & 1.3$\times$0.8 & \hphantom{$-$}32 & 59 & 340 & 6.8 & 16\hphantom{.41}(3)\\
G328.237$-$0.54 & 17 Dec 2005 & 15:57:58 & $-$53:59:23 & $-$94, $+$2 & 1.2$\times$0.8 & $-$16 & 52 &  \hphantom{0}98 & 5.5 & 5.0\hphantom{1}(1) \\
G329.469$+$0.50 & 16 Dec 2005 & 15:59:41 & $-$52:23:28 & $-$124, $-$28 & 1.5$\times$1.0 & \hphantom{$-$}14 & 61 & 410 & 6.9 & 5.3\hphantom{1}(1)\\
G329.031$-$0.19 & 17 Dec 2005 & 16:00:30 & $-$53:12:27 & $-$94, $+$2 & 1.2$\times$0.8 & $-$14 &  52 &  \hphantom{0}98 & 6.1 & 4.9\hphantom{1}(1)\\
G329.029$-$0.20 & 17 Dec 2005 & 16:00:32 & $-$53:12:50 & $-$94, $+$2 & 1.2$\times$0.9 & $-$12 & 52 &  \hphantom{0}98 & 6.1 & 4.9\hphantom{1}(1)\\ 
G329.066$-$0.30 & 17 Dec 2005 & 16:01:10 & $-$53:16:03 & $-$94, $+$2 & 1.2$\times$0.8 & $-$11 & 53 & 100 & 6.0 & 4.9\hphantom{1}(1) \\ 
G329.183$-$0.31 & 17 Dec 2005 & 16:01:47 & $-$53:11:44 & $-$94, $+$2 & 1.1$\times$0.9 & $-$3\hphantom{0} & 52 &  \hphantom{0}98 & 6.0 & 5\hphantom{.91}(1)\\
G331.442$-$0.18 & 16 Dec 2005 & 16:12:12 & $-$51:35:10 & $-$124, $-$28 & 1.8$\times$0.9 & \hphantom{$-$}16 & 51 & 350 & 7.3 & 7\hphantom{.31}(3)\\
G331.342$-$0.34 & 16 Dec 2005 & 16:12:26 & $-$51:46:17 & $-$124, $-$28 & 1.5$\times$0.9 & \hphantom{$-$}36 & 61& 410 & 7.1 & 5.3\hphantom{1}(1)\\
G332.295$-$0.09 & 17 Dec 2005 & 16:15:45 & $-$50:55:54 & $-$94, $+$2 & 1.1$\times$0.9 & \hphantom{$-$}34 & 52 &  \hphantom{0}98 & 7.9 & 5.3\hphantom{1}(8)\\
G332.604$-$0.16 & 17 Dec 2005 & 16:17:29 & $-$50:46:13 & $-$93, $+$3 & 1.2$\times$0.8 & \hphantom{0}$-$7 & 53 & 100 & 8.0 & 5.0\hphantom{1}(1)\\
G333.029$-$0.06 & 17 Dec 2005 & 16:18:57 & $-$50:23:54 & $-$93, $+$3 & 1.2$\times$0.9 & \hphantom{0}$-$9 & 52 &  \hphantom{0}98 & 8.4 & 5.0\hphantom{1}(1) \\ 
G333.315$+$0.10 & 17 Dec 2005 & 16:19:29 & $-$50:04:41 & $-$93, $+$3 & 1.2$\times$0.9 & \hphantom{0}$-$8 & 52 &  \hphantom{0}98 & 8.7 & 5.0\hphantom{1}(1)  \\
G333.184$-$0.09 & 16 Dec 2005 & 16:19:46 & $-$50:18:35 & $-$124, $-$28 & 1.7$\times$1.0 & \hphantom{$-$}29 & 57 & 390 & 8.5 & 5.0\hphantom{1}(5)\\
G333.163$-$0.10 & 25 Mar 2006 & 16:19:43 & $-$50:19:53 & $-$136, $-$40 & 1.3$\times$0.7 & \hphantom{$-$0}5& 61 &  \hphantom{0}76 & 4.9 & 5.3\hphantom{1}(1)\\
G333.23$-$0.05  & 16 Dec 2005 & 16:19:48 & $-$50:15:02 & $-$124, $-$28 & 1.8$\times$1.0 & \hphantom{$-$}20 & 54 & 370 & 8.6 & 5\hphantom{.01}(1)\\
G333.121$-$0.43 & 17 Dec 2005 & 16:21:00 & $-$50:35:52 & $-$94, $+$3 & 1.1$\times$0.9 & \hphantom{$-$}34 & 49 &  \hphantom{0}92 & 8.2 & 5.0\hphantom{1}(1) \\
G333.562$-$0.02 & 17 Dec 2005 & 16:21:09 & $-$49:59:48 & $-$93, $+$3 & 1.2$\times$0.9 & \hphantom{0}$-$3 & 52 &  \hphantom{0}98 & 8.8 & 5\hphantom{.01}(1)\\
G332.942$-$0.69 & 17 Dec 2005 & 16:21:19 & $-$50:54:10 & $-$94, $+$2 & 1.1$\times$0.9 & \hphantom{$-$}29 & 49 &  \hphantom{0}92 & 7.9 & 5.3\hphantom{1}(5) \\
G333.466$-$0.16 & 25 Mar 2006 & 16:21:20 & $-$50:09:49 & $-$105, $-$9  & 1.3$\times$0.7 & \hphantom{$-$0}8 & 60 &  \hphantom{0}75 & 4.6 & 5.2\hphantom{1}(1) \\
G332.963$-$0.68 & 25 Mar 2006 & 16:21:23 & $-$50:52:59 & $-$106, $-$10 & 1.3$\times$0.7 & \hphantom{$-$0}9 & 61 &  \hphantom{0}76 & 4.6 & 13.9\hphantom{1}(1)  \\
G333.130$-$0.56 & 25 Mar 2006 & 16:21:36 & $-$50:40:51 & $-$106, $-$10 & 1.4$\times$0.7 & \hphantom{$-$}12 & 61 &  \hphantom{0}76 & 4.5 & 5.2\hphantom{1}(1) \\
G335.060$-$0.42 & 26 Mar 2006 & 16:29:23 & $-$49:12:27 & $-$75, $+$22 & 1.5$\times$0.7 & \hphantom{$-$}19 & 57 &  \hphantom{0}71 & 3.7 & 5.4\hphantom{1}(1)  \\
G337.40$-$0.41  & 25 Mar 2006 & 16:38:50 & $-$47:28:18 & $-$103, $-$7  & 1.4$\times$0.7 & \hphantom{$-$0}9 & 61 &  \hphantom{0}76 & 3.8 & 5.1\hphantom{1}(1) \\
G338.92$+$0.56  & 16 Dec 2005 & 16:40:32 & $-$45:41:53 & $-$123, $-$27 & 2.0$\times$0.9 & \hphantom{$-$0}8& 50 & 340 & 13.6 & 5.7\hphantom{1}(3) \\
G338.92$-$0.06  & 25 Mar 2006 & 16:43:16 & $-$46:05:42 & $-$102, $-$6  & 1.4$\times$0.7 & \hphantom{$-$0}9& 61 &  \hphantom{0}76 & 4.8 & 13.8\hphantom{1}(1) \\
G339.62$-$0.12  & 26 Mar 2006 & 16:46:05 & $-$45:36:43 & $-$72, $+$24 & 1.6$\times$0.7 & \hphantom{$-$}17  & 57 &  \hphantom{0}71 & 5.2 & 5.6\hphantom{1}(1)  \\ 
G345.01$+$1.79  & 26 Mar 2006 & 16:56:49 & $-$40:14:20 & $-$69, $+$27 & 1.7$\times$0.7 & \hphantom{$-$}16 & 58 &  \hphantom{0}73 & 10.6 & 5.9\hphantom{1}(3) \\ 
G345.00$-$0.22  & 26 Mar 2006 & 17:05:10 & $-$41:29:04 & $-$69, $+$27 & 1.7$\times$0.7  & \hphantom{$-$}16 & 58 &  \hphantom{0}73 & 9.6 & 15.2\hphantom{1}(3) \\ 
G351.16$-$0.70  & 26 Mar 2006 & 17:19:57 & $-$35:57:46 & $-$66, $+$30 & 1.9$\times$0.7 & \hphantom{$-$}15 & 58 &  \hphantom{0}73 & 9.1 & 6.0\hphantom{1}(6) \\ 
G351.24$+$0.67  & 25 Mar 2006 & 17:20:16 & $-$35:54:58 & $-$24, $+$46 & 1.4$\times$1.0 & \hphantom{$-$}30 & 62 &  \hphantom{0}78 & 9.0 & 5.5\hphantom{1}(3)\\
G351.41$+$0.64  & 26 Mar 2006 & 17:20:53 & $-$35:47:02 & $-$66, $+$30 & 1.9$\times$0.7 & \hphantom{$-$}15 & 58 &  \hphantom{0}73 & 8.9 & 5.9\hphantom{1}(4)\\ 
G351.78$-$0.54  & 26 Mar 2006 & 17:26:43 & $-$36:09:18 & $-$66, $+$30 & 1.8$\times$0.7  & \hphantom{$-$}16 & 57 &   \hphantom{0}71 & 9.6 & 5\hphantom{.41}(1)\\ 
G0.54$-$0.85  & 25 Mar 2006 & 17:50:15 & $-$28:54:51 & $-$32, $+$64 & 1.7$\times$1.0 & \hphantom{$-$}23 &  60 &   \hphantom{0}75 & 8.3 & 14.5\hphantom{1}(5) \\
G9.621$+$0.196  & 26 Mar 2006 & 18:06:15 & $-$20:31:32 & $-$60, $+$37 & 3.0$\times$0.7 & \hphantom{$-$0}9 & 58 &  \hphantom{0}73 & 13.7 & 16\hphantom{.01}(1) \\ 
G18.34$+$1.77  & 25 Mar 2006 & 18:17:57 & $-$12:07:22 & $-$28, $+$68 & 3.7$\times$1.1  & \hphantom{$-$0}7 & 61 &  \hphantom{0}76 & 11.0 & 13\hphantom{.01}(3) \\
\hline
\end{tabular}
\end{table*}

\subsection{Notes on selected targets}
\label{NotesOnSources}

In this section we discuss the sources where the 9.9-GHz masers were detected along with
some non-detections. The latter are largely star-forming regions where an energetic 
interaction between outflow and molecular cloud has been reported in the literature,
and therefore conditions favouring detection of the 9.9-GHz maser could be expected.

\subsubsection{G301.14$-$0.23 (IRAS 12326$-$6245) non-detection}
This source is deeply embedded in a dense molecular cloud and harbours
one of the most energetic and massive bipolar molecular outflows known to date 
among objects of similar luminosity \citep{hen00}.  There is a weak ($\sim$1~Jy) class~II
methanol maser at 6.7~GHz with a peak velocity of $-$39.8~\kss \citep{cas95,cas09}. It is 
associated with one of the two compact
H{\sc ii} regions  present in the source \citep{wal98}. Within the uncertainty of the position
measurement this maser is co-located with the mainline OH maser, which has  features 
spread over unusually large velocity range from $-$64 to $-$35~\kss~\citep{cas87,cas98}.
Voronkov et al. (unpublished observations) have detected a class~I methanol maser at 25~GHz (J=5 transition of the 25-GHz series) peaking at $-$36.0~\kss in this source. No 
9.9-GHz maser emission has been detected in this source with the 1$\sigma$ flux density 
limit of 0.4~Jy.

\subsubsection{G305.21$+$0.21 possible detection}
The class II methanol maser at 6.7~GHz in this source was observed at high angular resolution by
\citet{nor93}, who found two centres of activity separated by approximately 22\arcsec, labelled G305A
and G305B following the notation of \citet{wal07} or
305.208$+$0.206 and 305.202$+$0.208 respectively according to \citet{cas09}.
There are two prominent H{\sc ii}
regions in the source located approximately 30\arcsec{ }to the south-east of G305A and 
15\arcsec{ }to the west of G305B, so none of the maser sites appear to be directly
associated with any continuum emission \citep{phi98,wal07}. According to \citet{wal07} the
region G305A is a hot core active in a number of molecular tracers, while the state of
G305B, which is associated with bright 8~$\mu$m emission is not fully clear.

This source was observed during the test 9.9-GHz survey without proper imaging. The spectrum
contains a hint (at 1.9$\sigma$) of maser emission near $-$42.5~\kss with flux density 
around 0.2~Jy. The regular
survey was not sensitive enough to confirm the detection. However, \citet{vor07}
reported a convincing detection of the 104~GHz methanol maser in the source at the same velocity.
The 104 and 9.9-GHz transitions belong to the same $(\mbox{J+1})_{-1}-\mbox{J}_{-2}$~E transition series
(J=10 and J=8, respectively) and are expected to show similar behaviour. Therefore,
it is likely that the 9.9-GHz maser is present in this source below the detection threshold of the survey.
The flux density of the 9.9-GHz emission is at least an order of magnitude lower than might have been expected on the basis of the only two
other sources known to have both 104 and 9.9-GHz masers. Both G343.12-0.06~\citep{vor06} and
W33-Met~\citep{vor07} have flux densities at 104-GHz comparable with the 9.9-GHz ones reported in
this paper (see Table~\ref{positive}). This most likely implies that markedly different physical conditions exist in
this particular source. One may speculate that this might be related to another peculiarity found
by \citet{wal07}: the high-mass star-forming region associated with G305B is bright in infrared, old according to
the chemistry, but has no H{\sc ii} region sufficiently developed to be detectable.
The expansion of an H{\sc ii} region could be hindered because of the dense environment and/or 
by continuing infall of the matter. 

\citet{vor05a} detected a bright class~I methanol maser at 25~GHz (J=5 transition
of the 25-GHz series) at $-$42.4~\ks, although without imaging and accurate flux density calibration. These
masers have been shown to originate at the same spatial location as the 9.9-GHz maser in at least one
source \citep{vor06}. An accurate position for the 25-GHz maser (J=4 transition with the peak velocity of $-$42.1~\kss
observed with the velocity resolution of 0.75~\ks) was determined by \citet{wal07}, but the reported flux
density of the maser is much lower, due to insufficient spectral resulution. The 25-GHz maser was found to
be 3\arcsec{ }to the east from the position of the 6.7~GHz maser  305.202$+$0.208 (G305B region). 
It is worth noting that
\citet{wal07} also detected 25-GHz emission (several transitions of the series) towards
two other locations in this source: near G305A and the south-eastern H{\sc ii} region. However, the radial
velocity and a clear maser appearance of the feature near G305B allows an unambiguous identification of the
suspected 9.9-GHz maser with this position, near the 6.7-GHz maser 305.202$+$0.208.


\subsubsection{G331.13$-$0.24 detection}
The 6.7-GHz masers in this source were mapped by \citet{phi98} and found to be located near
the edge of a 7$\times$10 arcsec$^2$ H{\sc ii} region (see also Fig.~\ref{maps}a). According
to \citet{goe04} this 6.7-GHz maser shows periodic flares with period around 500 days.
The OH maser in the source is coincident with the cluster of the 6.7-GHz maser spots within the
uncertainty of the absolute position measurement \citep{phi98,cas98}.
The 9.9-GHz maser is located close to
the cluster of class~II masers, but at a clearly distinct position. It is also seen projected onto the
H{\sc ii} region but located near  the edge. There is a protrusion in the image of the H{\sc ii}
region near the location of the 9.9-GHz maser (Fig.~\ref{maps}a) indicating a possible interaction
between expanding H{\sc ii} region and the surrounding material. The brightest 4.5-$\mu$m source 
within the boundaries of this H{\sc ii} region is located between the 6.7 and 9.9-GHz 
masers  (Fig.~\ref{maps}a). Two weaker 4.5-$\mu$m sources 
surrounded by an extended emission are located near the northern
tip of the H{\sc ii} region (as shown by an ellipse in Fig.~\ref{maps}a). This latter
object has an excess of the 4.5-$\mu$m emission and is listed as an EGO by 
\citet{cyg08} despite being only a few seconds of arc across.  The exact physical interpretation
of this source is uncertain at present. However, considering its location
it seems unlikely that this EGO could trace part of the shock front responsible for the 9.9-GHz maser.  

\citet{deb09} mapped the SiO (2-1) emission towards G331.13$-$0.24 and 
detected an elongated SiO source with velocity gradient
centred at the position of the 6.7-GHz masers with the axis roughly 
parallel with the East-West distribution of the 6.7-GHz  maser spots.
\citet{deb03} reported the detection of a single knot of
H$_2$ 2.12-$\mu$m emission. This is located at the eastern edge of the red-shifted SiO lobe 
(approximately 8\arcsec{ }from the 6.7-GHz maser). Both these facts reveal pronounced 
shock activity in the region. However, the present data do not allow us to
rule out the association of the 9.9-GHz maser with any certain structure seen in 
shock-tracing molecular emission  as 
the spatial resolution of the SiO data is rather coarse and comparable to the separation between 
the 9.9- and 6.7-GHz masers.  


\subsubsection{G338.92$-$0.06, G339.62$-$0.12, G9.62$+$0.20 non-detection}
These sources were included in the target list as periodically variable 6.7-GHz masers 
\citep{goe04}, although \citet{sly94} detected no class~I maser emission at 44~GHz 
towards either of them.  \citet{kur04} found weak 44-GHz maser emission
towards G9.62$+$0.20, which was below the detection threshold of \citet{sly94}.
Despite the fact that we observed all three sources  in a good
weather providing a 1-$\sigma$ flux density limit of around 
70-80~mJy (see Table~\ref{negative}), no emission at 9.9~GHz was detected.

\subsubsection{G343.12$-$0.06 (IRAS 16547-4247) detection}
The 9.9-GHz maser was discovered in this source during the pilot survey. It has been
investigated in detail in a separate paper \citep{vor06} and is included here for completeness.
The maser is associated with the brightest knot of the H$_2$ emission \citep{bro03} 
tracing an outflow
\citep[see Fig.~\ref{maps_assoc}a;][]{vor06}. This particular location is the only place
where the  25 and 104-GHz masers were detected in this source, in contrast to the 
widespread class~I methanol masers detected in a number of maser spots \citep{vor06}. 
\citet{gar03} first detected cm-wavelength 
continuum emission in the source comprising a brighter central object and two weaker 
satellites \citep[further resolved by][]{rod05}. 
The central object and the satellites were interpreted as thermal (free-free) emission 
from a radio jet and internal working surfaces of the jet, respectively. The 12-mm image
of these continuum sources from \citet{vor06} is shown in Fig.~\ref{maps_assoc}a 
along with the  2.12-$\mu$m molecular hydrogen image \citep{bro03} and the 9.9-GHz
maser position.
\citet{for89}
detected mainline OH masers at a position close to the central continuum source 
(Fig.~\ref{maps_assoc}a). No 6.7-GHz emission has been detected with an upper
limit of 0.3~Jy \citep{cas95,wal98}.

\subsubsection{G351.16$-$0.70, G351.24$+$0.67, G351.41$+$0.64 non-detection}
These are the positions within NGC~6334 complex, one of the most prominent
sites of massive-star formation. They correspond to sources V, IV and I/I(N), respectively 
\citep[see e.g.][and references therein]{mun07}. The massive twin cores NGC 6334I and I(N) 
have been studied for more than two decades and are believed to be at different evolutionary
stages \citep[see][and references therein]{beu08}.  These cores are well known to host unusual
methanol masers of both classes~\citep[e.g.,][]{men89, ell04,cra04}. There are class~II
methanol masers at 6.7~GHz associated with both cores~\citep{wal98}, but the maser in 
the core~I is one of the brightest known 6.7-GHz masers of more than 3000~Jy.
In contrast, the northern core, I(N), is remarkable in its class~I maser emission \citep[e.g.,][]{bro09}. 
Although both sites emit at 25 GHz, it is likely to be a maser in the I(N), while source~I
is responsible for quasi-thermal emission~\citep{men89,has90,beu05}. No 9.9-GHz maser emission has
been detected towards either of these positions with the 1$\sigma$ flux density limit of
70$-$80~mJy.


\subsubsection{W33-Met (G12.80$-$0.19) detection}

\begin{figure*}
\includegraphics[width=\linewidth]{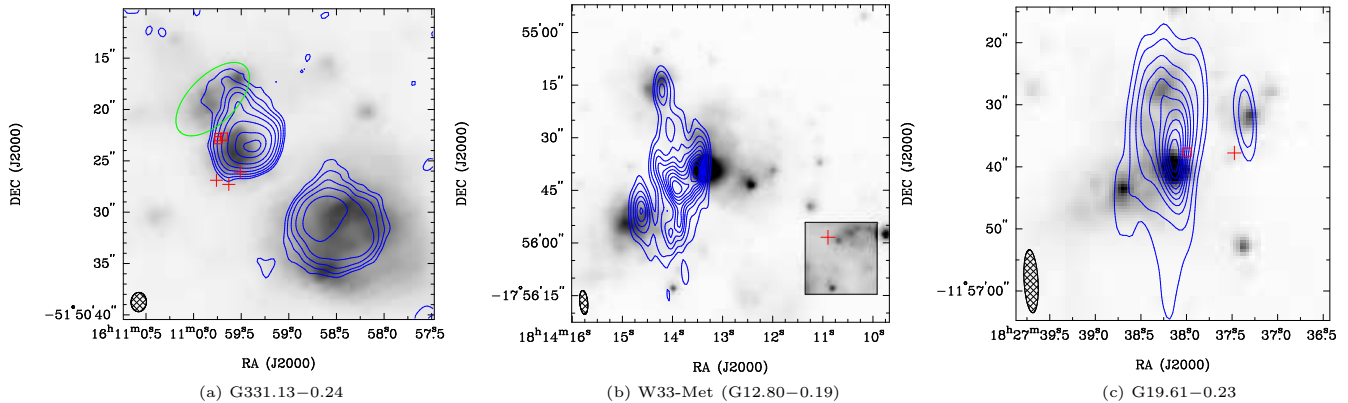}
\caption{Radio continuum emission at 8.6~GHz (contours) in the vicinity of masers detected 
at 9.9~GHz overlaid on top of the Spitzer IRAC 4.5-$\mu$m image (grayscale). The environment
of G343.12$-$0.06  is shown in Fig.~\protect\ref{maps_assoc}a and discussed in a separate 
paper \protect\citep{vor06}. 
Position of the 6.7-GHz and 9.9-GHz masers is shown by the open squares and crosses, respectively.
(a)~G331.13$-$0.24. The continuum image was obtained using the data of  \protect\citet{phi98},
to preserve high accuracy of the relative position of the 6.7-GHz maser with respect to the H{\sc ii} region.   Contours are 1, 2, 4, 8, 16, 32, 64 and 90\% of 43~mJy~beam$^{-1}$.
The crosses represent 3 epochs of the measurement at 9.9~GHz, 
rather than individual maser spots. The ellipse represents the location of an EGO \protect\citep{cyg08}. 
(b)~W33-Met (G12.80$-$0.19). The 4.5-$\mu$m flux density has been multiplied by 10
inside the square in the bottom right corner of the image. Contours are
5, 10, 20, 30, 40, 50, 60, 70, 80 and 90\% of 1.8~Jy~beam$^{-1}$. 
(c)~G19.61$-$0.23.
Contours are 5, 10, 20, 30, 40, 50, 60, 70, 80 and 90\% of 0.6~Jy~beam$^{-1}$.}
\label{maps}
\end{figure*}

The W33 region, which hosts a cluster of young massive stars, is renowned for its strong 
radio continuum 
emission and a complex molecular environment \citep{gos78,gol83,ho86}. The first detection 
of a methanol maser towards this source was reported by \citet{men86}. They observed
a number of class~I maser transitions of the 25-GHz series  and revealed a positional
offset of the masers  with respect to the dominant continuum source. The 44-GHz methanol 
masers were later found in the vicinity of this offset position by \citet{has90}, who introduced the notation
W33-Met to distinguish methanol maser site from the main continuum source in the region.
Before we started the project, this was the only source known to have a 9.9-GHz maser
\citep{sly93}.  Although our measurement gives the peak flux density as 4.3~Jy, \citet{sly93}
reported just 0.8~Jy. Partially, this is the result of the 2.5 times lower spectral resolution in the \citet{sly93}
data, as the line is quite narrow (Fig.~\ref{spectra}). However, the spectral resolution alone cannot 
account for the more than factor of 4 difference in the flux density. It is not clear how accurate the
absolute flux scale calibration was for the \citet{sly93} single dish data, especially given the 
strong cm-wavelength continuum in the source.  Therefore, we do not draw any conclusions 
from this discrepancy, although temporal variability cannot be excluded. \citet{sly93}
reported the peak velocity of the maser at 33.5~\ks, which is notably different from 
32.72~\kss obtained in our measurement (Table~\ref{positive}) even taking into consideration
the spectral resolution achieved in both experiments.  However, this discrepancy  can be explained by the
less accurate value of the rest frequency used by \citet{sly93} to calculate the radial velocities. 
The difference in adopted rest frequencies accounts for 0.85~\kss of the velocity offset, bringing the
radial velocity into agreement within the spectral resolution. The radial velocity of the 9.9-GHz maser
is also in agreement with the velocity of the 25-GHz masers reported by \citet{men86}, provided the rest frequency adopted by \citet{men86} is more accurate than the spectral resolution of their measurement.  

The continuum source has a complex structure reflecting the presence of multiple
young stars (Fig.\ref{maps}b).  The morphology is in general agreement with the earlier results of
\citet{ho86}, but the 1.5-GHz emission (20cm) is extended more to the west than the
emission at higher frequencies. This is most obvious if one compares the association between the (1,1) inversion 
transition of ammonia and the rim of the 1.5-GHz emission depicted by \citet{ket89} in their Fig.~2
with Fig.~\ref{maps_assoc}b in this paper, where the same ammonia distribution from
\citet{ket89} is overlaid on top of the 8.6-GHz continuum from our observations.
Fig.~\ref{maps_assoc}b clearly shows a gap of around 15\arcsec{ }between the 8.6-GHz continuum
source and the ridge of ammonia emission.  As concluded by \citet{ho86}, the smooth emission
filling the gap is most likely resolved at higher frequencies.  According to the interpretation of 
\citet{ket89} the ridge of ammonia emission is due to shocks caused by interaction of the ionised gas
with the neutral component.  The 9.9-GHz maser is located near
the western clump of the ammonia emission (Fig.~\ref{maps_assoc}b). This position is in agreement 
with the position of the 25-GHz masers reported by \citet{men86} within the uncertainty of their measurement.

The Spitzer IRAC image revealed a number of 4.5-$\mu$m sources associated with
some of the continuum peaks (see Fig.\ref{maps}b). There is an extended 4.5-$\mu$m 
emission located largely near the western edge of the 8.6-GHz continuum source in the area where
the continuum emission has been detected at 1.5~GHz only. The inspection of other IRAC bands
revealed no significant excess of the 4.5-$\mu$m emission, which is in agreement with
this object not being listed as an EGO by \citet{cyg08}. However, it is worth noting that
there is a faint compact 4.5-$\mu$m source located just 3\arcsec{ }to the west of the 9.9-GHz maser position. This infrared source is detected in the 4.5-$\mu$m IRAC band only. 


Although \citet{men91} reported the detection of the 6.7-GHz maser in W33-Met, a comparison of the velocity range
with the source G12.91$-$0.26 reported by \citet{cas95} suggests possible confusion. We processed archival
ATCA data (project code C415) using the standard reduction technique and found two 6.7-GHz masers
G12.68$-$0.18 ($\alpha_{2000}=18^h13^m$54\fs74$\pm$0\fs01, $\delta_{2000}=-18$\degr01\arcmin46\farcs1$\pm$0\farcs1) and
G12.91$-$0.26 ($\alpha_{2000}=18^h14^m$39\fs52$\pm$0\fs02, $\delta_{2000}=-17$\degr51\arcmin59\farcs1$\pm$0\farcs2),
situated 7\farcm0 and 7\farcm9 away from the 9.9-GHz methanol maser position, respectively. The 
1$\sigma$ rms noise level at the position of the 9.9-GHz maser at 6.7~GHz is 0.1~Jy. \citet{cas09}
provided an independent position measurement of the confusing 6.7-GHz masers which agrees 
well with the values given above.

\subsubsection{G18.34$+$1.77 (IRAS 18151$-$1208) non-detection}
Although the widespread class~I maser transitions (e.g. at 44~GHz) are not atypical in this source, 
it has one of the brightest  25-GHz methanol masers found outside Orion and NGC~6334 
(the J=5 maser peaks at $+$29.7~\kss and has a flux density exceeding 15~Jy, Voronkov et al., unpublished observations). 
\citet{dav04} detected two H$_2$ outflows in this source which could be related to the 25-GHz masers.
High angular resolution observations are required to confirm this. No 9.9-GHz maser emission has
been detected in this source with the 1$\sigma$ flux density limit of 76~mJy, which implies that
the flux density of the 25-GHz maser exceeds that of the 9.9-GHz maser by at least two orders of
magnitude.

\subsubsection{G19.61$-$0.23 detection}
This region  hosts a cluster of young high-mass stars and is somewhat similar in appearance 
to W33-Met at cm-wavelength (Fig.~\ref{maps}b and c), albeit with a smaller angular
separation between components. \citet{fur05} identified nine ultra-compact
H{\sc ii} regions and a hot core squeezed between two of them designated A and C. 
The 8.6-GHz continuum image
(Fig.~\ref{maps}c) obtained in our survey qualitatively agrees with the results of \citet{fur05}, 
although the elongated beam does not allow us to disentangle close components easily.   
A number of continuum peaks coincide with the 4.5-$\mu$m emission from   
the Spitzer IRAC image (see Fig.\ref{maps}c). The brightest 4.5-$\mu$m source is 
associated with the most prominent H{\sc ii} region \citep[labelled A in][]{fur05}. 
This source is not listed as an EGO by \citet{cyg08}.
 
\begin{figure*}
\includegraphics[width=\linewidth]{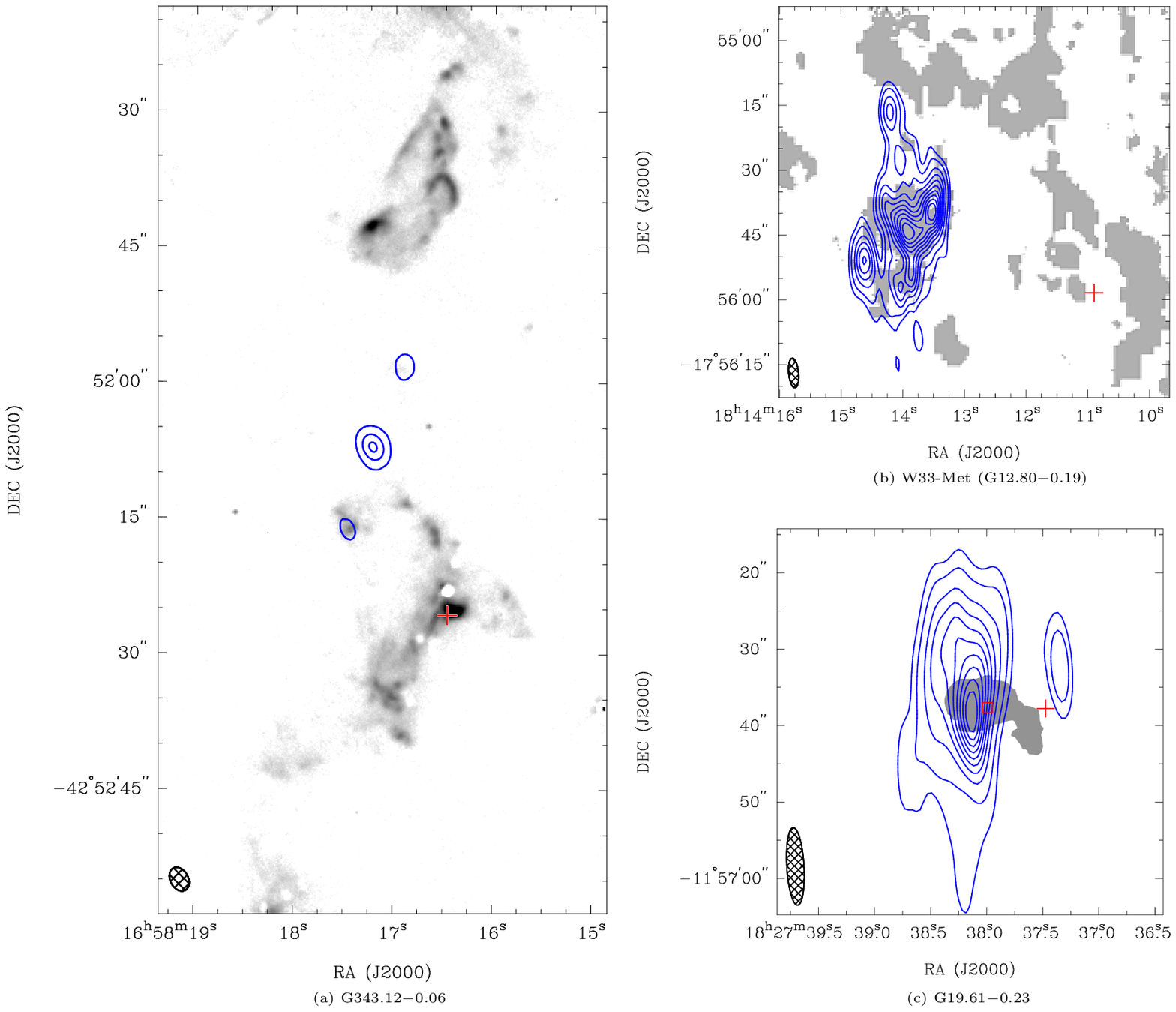}
\caption{Association of the 9.9-GHz masers with outflow and hot molecular 
gas in the vicinity of H{\sc ii} regions. Position of 
the 9.9-GHz masers is shown by crosses and the 6.7-GHz maser in just G19.61$-$0.23 by
an open square.
(a)~G343.12$-$0.06. The maser position is overlaid on top of the H$_2$ 2.12-$\mu$m 
emission \protect\citep{bro03} and the 12mm continuum \protect\citep{vor06} showing
an association with outflow. Contours are 15, 50 and 90\% of 13~mJy~beam$^{-1}$.
(b)~W33-Met (G12.80$-$0.19). The grayscale shows the (1,1) ammonia 
emission  \protect\citep{ket89}. Contours show the 8.6-GHz continuum as in Fig.\ref{maps}b, 
(c)~G19.61$-$0.23. The grayscale represents the (2,2) ammonia
emission from \protect\citet{gar98}. The northern molecular clump seen in absorption
is not shown in this figure. Contours show the 8.6-GHz continuum as in Fig.\ref{maps}c,}
\label{maps_assoc}
\end{figure*}

The hot core is 
responsible for emission in a number of molecular species \citep{fur05,wu09} and is associated
with the brightest ammonia clump (named middle clump) found by \citet{gar98}. 
\citet{gar98} identified two more ammonia clumps: the northern clump seen in absorption and
the elongated south-western clump seen in emission. The relative position of the 8.6-GHz
continuum, ammonia emission \citep[from][absorption source is not shown]{gar98} and the 9.9-GHz
maser is shown in Fig.~\ref{maps_assoc}c. The 9.9-GHz maser is located near the 
elongated south-western clump, which \citet{gar98} interpreted as arising from the compressed
and heated gas behind the shock front driven by the expansion of the largest H{\sc ii} region
in the source.
\citet{kur04} observed this source at 44~GHz and revealed two maser spots.
The brightest 44-GHz spot is co-located both in position and velocity with the 9.9-GHz maser within the
combined error of the measurements. The weaker feature peaks 
at $-$46.5~\ks. It corresponds to a maser spot  located 7\farcs9 to the 
North-East from the brightest spot (and the 9.9-GHz maser spot) roughly along the ammonia 
filament.

\citet{cas95} detected a weak 6.7-GHz methanol maser with a peak flux density of  0.4~Jy in this source. This maser is located 
7\farcs6 away from the position of the 9.9-GHz maser,
at $\alpha_{2000}=18^h27^m$37\fs993$\pm$0\fs006,
$\delta_{2000}=-11$\degr56\arcmin37\farcs6$\pm$0\farcs4 (Caswell, unpublished observations).
The position suggests an association of this maser with the hot core and the middle ammonia 
clump (see Fig.~\ref{maps_assoc}c). \citet{gar85} found a number of OH maser spots clustered
near the 6.7-GHz maser position and also associated with the middle ammonia 
clump \citep[see also][]{for89}.

\section{Discussion}

\subsection{Temporal variability of G331.13$-$0.24}


Inspection of the 3 epochs listed in Table~\ref{positive} suggests possible temporal variability of 
the 9.9-GHz maser in G331.13$-$0.24, with the caveat that the only measurement which stands out 
significantly was taken under imperfect weather conditions. The variability of the class~I methanol
masers is not well understood at the moment. However, the class~II maser at 6.7~GHz 
in this source is known to be somewhat exceptional. This is one of the seven periodically variable 
masers known to date \citep{goe04}, which has the longest period in the sample of around 500 days. 
The coexistence of 6.7 and 9.9-GHz masers in this source is  a unique situation with great potential 
to shed light on the nature of periodic flares, because these two
transitions are predicted to react in opposite ways to the changes in pumping.
The pumping mechanisms of these two maser transitions are in conflict as they belong to different classes.
Hence, if the flares occur due to a pumping variation  of any nature, the intensity
variations in the two transitions are expected
to anti-correlate. From the other hand, both masers are seen projected towards the 
same H{\sc ii} region. This means that this H{\sc ii} region provides the seed radiation which 
both masers amplify. If the variations are due to changes in the seed radiation, intensity variations 
in the two transitions will correlate.  Thus simple monitoring of the 9.9-GHz
maser may unravel a vital aspect of the flare mechanism. We have started such a project  
and will report the results elsewhere.

The main complication arises from the
fact that class~I and class~II masers are not co-located. In this and some other periodically 
variable sources, flares of the different 6.7-GHz spectral features, which are presumably
also separated in space, are delayed with respect to each other, sometimes by up to a few 
weeks \citep{goe04}.  Although the velocity of the 9.9-GHz maser is close to the velocity of one of 
the groups of the 6.7~GHz maser lines, the angular separation is an order of magnitude greater than the
size of the 6.7-GHz cluster.  Therefore, a longer delay of the order of months is quite possible.
The 9.9-GHz flux measurements presented in this paper all correspond to the descending part of the
6.7-GHz light curve, approximately 20, 4 and 10 weeks after the  maximum 
(Goedhart et al., unpublished recent monitoring data), respectively for each epoch 
listed in Table~\ref{positive}.  Although this is consistent with the steady fall of the 9.9-GHz flux 
density along with the 6.7-GHz flux density, a longer time series is required 
to obtain a definitive answer.
The long period of around 500 days as well as the non-sinusoidal shape of the light
curve \citep[see][]{goe04} will help to distinguish between a possible delay and 
anti-correlation in this particular source.

Monitoring of the masers has a number of  advantages over monitoring the continuum emission directly. 
Firstly, the masers amplify the continuum and so essentially undetectable
variations of the continuum flux can cause significant variations in the maser intensity.  Secondly, the
continuum source has a complex shape in G331.13$-$0.24 making it very difficult to undertake  accurate
flux density monitoring as it is hard to reproduce exactly the same uv-coverage in all epochs.
In contrast, each spectral feature of the methanol maser corresponds to a point source for the
baselines of the order of a few kilometres long, making flux monitoring a much easier task. It is worth 
noting that there is an agreement to better than 2\% (which is likely to be the limit of the absolute 
flux density calibration) between the flux density of the H{\sc ii} region measured
by \citet{phi98} and that obtained in our study.

\subsection{Coexistence with class II methanol masers at 6.7 GHz}
\label{AssociationClassII}

As was mentioned above, our target list comprised largely known class I methanol masers.
The majority of these class~I masers were found towards known 6.7-GHz masers (class~II),
as they are often located within the beam of the single dish used for class~I measurements 
\citep[e.g.][]{sly94,ell05}. The remaining
sources were either well known star-forming regions (e.g. positions within the NGC6334 complex)
or were found towards known H$_2$O masers \citep[e.g.][]{bac90}, H{\sc ii} regions
or main-line OH masers \citep[e.g.][]{sly94}. Despite being selected by other criteria
many of the sources from the second group also have a 6.7-GHz methanol maser
in the vicinity. Therefore, only one target in our sample, W33-Met,
has a class~I methanol maser without a 6.7-GHz counterpart. Another such source is
G343.12$-$0.06 \citep{vor06}.

In total, out of four 9.9-GHz methanol masers currently known, two do not have a 6.7-GHz counterpart detected, and
one source, G19.61$-$0.23, has a very weak 6.7-GHz maser. The fourth source, G331.13$-$0.24,
has both masers projected onto an H{\sc ii} region, a situation favouring detection of low gain masers.
These results suggest that the 9.9-GHz masers tend to be associated with sources with
class~I masers without strong class~II counterparts. A proper statistical
test is not yet possible due to a small number of known 9.9-GHz masers.
However, a number of previous studies have already investigated relations between the widespread class I masers
(e.g. at 44 or 95~GHz) and the class~II masers at 6.7~GHz \citep[e.g.][]{sly99, ell05}.
\citet{sly99} were motivated by the conflicting pumping mechanisms of these masers,
although the present data suggest that the class~I and class~II masers are rarely seen
at the same location at high spatial resolution \citep[e.g.][]{kur04,cyg09}. The exceptions are
probably due to the projection effect, although unusual pumping under common 
conditions \citep[e.g.][]{vor05b} cannot be
excluded. Therefore, any tendency (e.g. for bright masers of one class to avoid the bright masers of another class),
if present, is likely to be related to the general trends in the distribution of the physical parameters
affecting the pumping across the star-forming region, rather than the conditions at any single location
in the source. These trends can be idiosyncratic to some extent due to the complex structure of high-mass star-forming
regions (e.g. one region could be more embedded in the parent molecular cloud than another). However, they can also
change during the course of evolution, for example when an outflow turns on or an H{\sc ii} region starts to expand.
It is worth noting that using a statistically complete sample of sources, \citet{ell05} found no evidence
for the original anti-correlation between flux densities reported by \citet{sly99}. However, the 9.9-GHz masers
require a more restrictive range of physical conditions (e.g temperature and density) to appear than the wide-spread
class~I methanol masers at 44 and 95~GHz. It is therefore possible that the anti-correlation 
with the 6.7-GHz masers
is more pronounced for these 9.9-GHz masers than for the widespread 
44 and 95-GHz ones.

\subsection{Association with expanding H{\sc ii} region}
\label{AssociationHII}

Out of 48 sources searched
for the 9.9 GHz emission in this study (note that G329.031-0.19 and G329.029-0.20 are counted as a
single source), 35 sources have 8.6-GHz continuum sources, presumably H{\sc ii} regions, 
within the primary beam (half power width is about 5 arcmin) and 13 sources do 
not (1$\sigma$ rms noise level not corrected for decorrelation and primary beam attenuation is about 0.3~mJy). Three of these
continuum sources correspond to periodically variable 6.7-GHz masers and two of them do not 
have class~I 
methanol masers detected so far in any transition. Even allowing for a few confused cases due 
to a large beam (most known class~I masers have single-dish positions), our target
sample, comprising largely known class~I masers at 44 and 95~GHz, appears to be biased to some extent towards the star-formation sites
with H{\sc ii} regions. The present data do not yet allow us to establish reliably whether 
the common occurrence for class~I masers to have an H{\sc ii} region in the vicinity has physical
significance or is just a selection effect.
In contrast, it is well established that only a small fraction of class~II methanol masers at 
6.7~GHz are associated with detectable H{\sc ii} regions \citep[e.g.,][]{wal98}. 

  
All four currently known 9.9-GHz maser sources have a cm-wavelength continuum detected
in the vicinity of the maser spot. Three masers found in the regular survey are 
located in close proximity to well-developed H{\sc ii} regions (Fig.~\ref{maps}). 
While the remaining source, G343.12$-$0.06, is renowned for thermal emission from the jet driving an
outflow \citep[Fig.~\protect\ref{maps_assoc}a;][]{bro03,vor06}. As mentioned in 
section~\ref{NotesOnSources}, in some cases the molecular emission could arise due to interaction between ionised and neutral material. Such an interpretation has been suggested on the
basis of kinematics and morphology  for at least two sources, 
W33-Met and G19.61$-$0.23 \citep{ket89,gar98}. The expansion of an H{\sc ii} region drives 
shocks into the surrounding molecular cloud heating and compressing the gas. It is this change in the 
physical conditions which is most likely traced by the ammonia emission (Fig.~\ref{maps_assoc}bc). 
In both  W33-Met and G19.61$-$0.23, the 9.9-GHz maser is located just beyond the 
edge of the ammonia 
source, further away from  the H{\sc ii} region which is believed to be undergoing expansion.  The separation
between the maser and the nearest ammonia emission is of the order of a few seconds of arc. Although the magnitude of this separation could probably be explained by the combined accuracy of the absolute 
positions, the picture agrees well with the basic expectation that a more fragile molecule like methanol should be located predominantly further away from the exciting source than ammonia. 

\citet{kir09} modelled 
the expansion of an H{\sc ii} region surrounded by molecular gas and demonstrated formation of a transition 
layer characterised by high abundance of molecules in the gas phase. The location and thickness of this
transition layer vary from molecule to molecule and also depend on the initial temperature and density of
the quiescent gas. Although the methanol chemistry was not modelled by \citet{kir09}, the physical
conditions in the transition layer, such as temperature and density, simulated in these models 
could clearly be in the range suitable
for the formation of class~I masers  \citep[e.g.,][]{vor05b}. 

The geometry of G331.13$-$0.24 and the data on the thermal molecular environment available for this
 source do not allow us to make conclusions about the nature of the 9.9-GHz maser. 
As previously mentioned in section~\ref{NotesOnSources}, there is an SiO emission (presumably
tracing shocks) detected in this source
 \citep{deb09} at a position which does not allow us to completely rule out the association with the 
maser.  On the other hand, the 9.9-GHz maser is located near the protrusion on the southern
side of the H{\sc ii} region (see Fig.~\ref{maps}a) which may reflect the presence of the same type of interaction between the ionised and neutral material as described above. Although the remaining
source G343.12$-$0.06 is a clear case where the 9.9-GHz maser is associated with an 
outflow~\citep[Fig.~\protect\ref{maps_assoc}a;][]{vor06}, these considerations  show that the outflow scenario may not be the only option. 

This idea should equally apply to other class~I masers, including
36 and 44-GHz ones, which are more ubiquitous than the 9.9-GHz masers. Therefore, future 
interferometric surveys of class~I masers, e.g. a follow up of the MMB survey 
\citep[briefly mentioned in ][]{cas10}, have a great potential to test this hypothesis. It is worth mentioning that \citet{kog98} observed W33-Met at 44~GHz. 
However, phase referencing was not used in their experiment making the search for possible associations difficult due to a lack of absolute positions. 

 
\subsection{Evolutionary stages}

The question whether different masers trace distinct evolutionary stages of
high-mass star formation has been investigated for more than two decades
\citep[e.g.,][]{for89,cod00,ell05,ell06, vor06, bre10}. The majority of class~II methanol masers
at 6.7~GHz are associated with millimetre and submillimetre sources \citep{wal03,hil05}, 
while only a small fraction shows association with detectable H{\sc ii} regions 
\citep{phi98,wal98}. This leaves no doubt that they trace a very early stage of the (proto)stellar
evolution preceding the formation of an H{\sc ii} region, and fade out shortly after
an H{\sc ii} region forms.  \citet{bre10} find  that the other widespread 
class~II methanol masers at 12.2~GHz most likely lag behind the 6.7-GHz masers.
The latter seem to largely overlap with the H$_2$O masers \citep{szy05} whereas
the OH masers are usually considered to be a signature of a 
somewhat more evolved stage, since they show a greater overlap with the 
ultracompact H{\sc ii} regions than is typical for class~II methanol masers \citep{for89,cas97}. 

However,  it is still poorly understood as to where the class~I methanol masers fit into this picture.
The main obstacle is the lack of unbiased surveys of the class~I masers, which cover a substantial
part of the Galaxy. As discussed above, the majority of currently known class~I 
methanol masers have been found in surveys targeted towards class~II methanol  masers.
Therefore, the sample of currently known class~I masers is strongly biased towards the evolutionary 
stage traced by the 6.7-GHz masers. Without a widespread and bright low-frequency class~I 
methanol maser transition (similar to the class~II maser transition at 6.7~GHz or mainline OH)
a blind survey is very time consuming. Therefore, \citet{pra08} adopted an interesting strategy
to cover several known molecular clouds in their, otherwise untargeted, search for class~I 
methanol masers at 36 and 44~GHz.  Comparison with the presence of H{\sc ii} regions and infrared 
sources, led \citet{pra08} to suggest that the 36-GHz masers may appear earlier than the 44-GHz masers
during the evolution of a star-forming region. However, these data do not allow  
interrelations between evolutionary stages where the class~I and class~II methanol masers
are present to be established. \citet{ell06} investigated this question using the infrared properties of a subsample of 
methanol masers associated with GLIMPSE (Galactic Legacy Infrared Mid-Plane
Survey Extraordinaire) catalogue point sources. It was found that the sources with a class~I
maser seem to have redder GLIMPSE colours than those without it. With the caveat that
the small number statistics made any firm conclusion impossible, \citet{ell06} suggested that the
class~I methanol masers may signpost an earlier stage of high-mass star formation than
the class~II masers.  

These and other considerations led \citet{ell07} to propose a qualitative evolutionary scheme
for the different maser transitions, which has subsequently been further refined by \citet{bre10} in their Figure 6. In this sequence the class~I masers represent the
first signatures of high-mass star-formation and fade out within approximately 2$\times$10$^4$ years after the onset of the 6.7-GHz maser emission. Based on the detailed study of G343.12$-$0.06, \citet{vor06} pointed out that the suggestion that class~I methanol masers precede class~II is inconsistent with the presence of OH masers and the lack of the 6.7-GHz maser emission in this source. OH maser
emission in the absence of 6.7-GHz masers is more consistent with 
an evolved stage of star-formation \citep[see e.g.][]{cas97}.  

It is important to keep in mind the major assumptions which underlie the evolutionary 
timeline put forward by \citet{ell07} and 
\citet{bre10}.  In particular, that each major maser species arises only once during the evolution of a particular star formation region, and that all the maser species are associated with a single astrophysical object.  For regions where the positions of the different maser species are known to an accuracy of an arcsecond or better this latter assumption can be proven or disproved for most transitions; however, the larger spatial (and angular) scale over which class~I methanol masers are often observed makes it much less clear whether the emission is due to a single object 
\citep[see, e.g.,][]{kog98,kur04,vor06,cyg09}.  
The degree of overlap of the class~I and class~II methanol masers on the timeline of \citet{ell07} and
\citet{bre10} is based on the rate of detection of class~I methanol masers towards class~II masers
observed by 
\citet{ell05}.  However, the idea that class~I methanol masers  precede the class~II was based
purely on the circumstantial evidence that the former are associated with outflows, which are generally considered to be signposts of young objects, and the suggestion of \cite{ell06} that the class~I masers may be associated with redder GLIMPSE point sources.  As discussed earlier, all the 9.9-GHz masers found in this study are located near H{\sc ii} regions (see section~\ref{AssociationHII}),
all but W33-Met have OH masers detected (see section~\ref{NotesOnSources}; no sensitive OH 
data seem to be available for W33-Met), and one source, W33-Met, does not have a  6.7-GHz 
maser detected (see section~\ref{AssociationClassII}). Despite the small number of sources, 
these similarities encourage 
us to revisit the question of where class~I methanol masers fit into a maser-based evolutionary sequence.

Contrary to the sequence presented by \citet{bre10} it is clear that class~I masers do overlap in time with
OH masers.  The 9.9-GHz maser detections reported in this paper reinforce the conclusion, although
the class~I masers in other transitions were already known for all of these sources 
\citep[e.g.,][]{bac90,sly94,lie96}. Despite the
biased target selection in the majority of class~I maser surveys, a few sources such as
W33-Met  (see section~\ref{NotesOnSources}) and G343.12$-$0.06 \citep{vor06}, are known to have 
class~I masers without a class~II counterpart at 6.7-GHz (see also section~\ref{AssociationClassII}).  
This suggests that it is more appropriate to place the class~I methanol masers as partly overlapping, but largely post-dating the evolutionary phase associated with class~II methanol masers.  As a part of the follow up program based on the MMB 
untargeted survey briefly mentioned by \citet{cas10}, class~I methanol
masers at 44~GHz were searched for towards a number of OH masers which have no detected 
6.7-GHz masers. The full results of this work will be reported elsewhere. However,  the preliminary analysis 
indicate a detection rate greater than 50\%.  This supports the hypothesis on the lifetime of class~I masers
given above and moreover suggests that the result is not unique for the 9.9-GHz masers and
also applies to the widespread class~I methanol masers (e.g. that at 44~GHz).
 
The class~I masers are formed in moderately dense gas separated from strong infrared 
sources \citep[for discussion see, e.g.,][]{vor06}. These masers have radial velocities close to that of a 
parent molecular cloud which suggests, together with morphology, an association with the interface region. Weak shocks  interacting with the molecular cloud dramatically change the chemical
composition of the medium by releasing methanol and other complex molecules from the
dust grain mantles into the gas phase. This increase of methanol abundance is observed in the
shock-processed regions related to outflows \citep[e.g.][]{gib98}. 
However, the same effect is also expected if other mechanisms, not just outflows, drive shocks into the parent molecular cloud, provided the shocks are 
weak enough and do not dissociate  methanol molecules. 
In particular, as discussed in section~\ref{AssociationHII}, class~I methanol masers could be 
associated with expanding H{\sc ii} regions as well as with outflows. The presence of an H{\sc ii} 
region implies that the star formation in a particular site is quite evolved. On the other hand,
an ample amount of molecular material interacting with an outflow or expanding H{\sc ii} region
is required to produce a detectable maser. Therefore, there could be a bias towards deeply
embedded regions of star formation. The amount of quiescent material remaining in a star-forming region showing maser activity depends not only on the age, but also on the initial conditions which 
may vary from one site to another.

It it important to recall that high-mass star formation usually occurs in a crowded environment
where sequential or triggered star formation may take place. G19.61$-$0.23 and W33-Met are good examples of clustered star formation.  For example, \citet{ho86} suggested that the interaction 
of ionised gas from an earlier epoch of star formation may have triggered the formation of a second
generation stars in W33-Met through the radiation-driven implosion mechanism. Therefore,
in some cases it may be impossible to attribute a class~I maser spot to a particular YSO, especially
taking into account that the angular offset of this maser may exceed the separation between YSOs
in the cluster. In addition, the individual constituents of the cluster may have significantly different
ages. For example, one YSO in the cluster may have an associated 6.7-GHz maser, while another
could be associated with OH masers and a well developed H{\sc ii} region. The fact that the
class~I methanol masers can be associated with both outflows and expanding H{\sc ii} regions,
which presumably appear at different evolutionary stages, further complicates the situation. At this
stage the number of sufficiently simple sources studied in detail is quite limited hindering the statistical
analysis. The future interferometric surveys of the 44- and 36-GHz methanol masers (widespread class~I maser transitions) will be vital for refining the evolutionary sequence for masers.

\subsection{Association with the EGOs}

\citet{che09} found that approximately two-thirds of the EGOs from the list of \citet{cyg08}, which 
have been observed at either 95 or 44~GHz, have a class~I methanol maser detected within 1\arcmin. 
Based on this result, they made a comment that EGOs could be good targets for future class~I
maser searches. Such a survey of the 44-GHz masers carried out by \citet{cyg09} indeed had a 
high detection rate (90\%, 17 detections out of 19 EGOs observed). However, two out of four
9.9-GHz  masers reported in this study 
are not associated with an EGO (see section~\ref{NotesOnSources}). 
The physical nature of EGOs requires
a better understanding \citep{cyg09}. One may speculate that the outflows are more likely to 
produce an EGO
when they interact with the parent molecular cloud, than if shocks are caused by an expanding 
H{\sc ii} region.  It is worth mentioning, that the survey of \citet{cyg09} discovered an interesting
class~I methanol maser G49.27$-$0.34. No 6.7-GHz maser has been detected in this 
source~\citep{cyg09}, while \citet{meh94} detected smooth radio continuum from 
presumably a well evolved H{\sc ii} region (seen at 20cm only). 
 
\section{Conclusions}
\begin{enumerate}
\item Two new class~I methanol masers at 9.9-GHz were found in addition to 
two other sources already reported in the literature. The absolute positions with arcsecond accuracy
are summarised in Table~\ref{positive} for all four 9.9-GHz masers known to date. Based on the
trial observations, we also suspect that another 9.9-GHz maser may exist in G305.21$+$0.21 at
a flux density below the detection threshold of our survey.
\item We suggest that some class~I methanol masers  may be associated with shocks driven
into molecular cloud by an expanding H{\sc ii} region. This is an alternative scenario to the
association with outflows which has been proved unambiguously in a number of other 
cases \citep[e.g.,][]{kur04,vor06}. This new scenario applies also to the class~I methanol 
masers in other transitions, e.g. to the widespread masers at 36 and 44~GHz. It does not appear
to be an exclusive mechanism to generate the 9.9-GHz masers, although could be responsible for
3 out of 4 known 9.9-GHz masers. 
\item The evolutionary stage with the class~I maser activity is likely to outlast the stage when
the 6.7-GHz methanol masers are present and overlap significantly in time with the stage when the OH
masers are active. We expect that the class~I masers in one of the widespread maser transitions
(e.g. at 44~GHz) will be detected towards a significant number of OH maser sources which do not have 
a class~II methanol maser at 6.7~GHz.  Such sources were rarely observed in the class~I maser surveys currently available in the literature.
\item A 9.9-GHz methanol maser was found in G331.13$-$0.24 which also hosts one 
of the few known periodically variable class~II methanol masers at 6.7~GHz. This rare combination
makes this source an attractive target for a long-term monitoring program, which has a 
high potential to shed light on the periodic variability of masers.
 
\end{enumerate}

\section*{Acknowledgments}
The Australia Telescope is funded by the
Commonwealth of Australia for operation as a National Facility managed
by CSIRO. SPE thanks the Alexander-von-Humboldt-Stiftung for an Experienced Researcher Fellowship which has helped support this research.  
AMS was financially supported by the Russian Foundation for Basic
Research (grant 09-02-97019-a) and the Russian federal program
"Scientific and scientific-pedagogical  personnel of innovational
Russia" (contracts N~02.740.11.0247 from 07.07.2009 and N~540 from
05.08.2009). This research has made use of NASA's
Astrophysics Data System Abstract Service. This research has made use of data products from the
GLIMPSE survey, which is a legacy science program of the {\em Spitzer
Space Telescope}, funded by the National Aeronautics and Space
Administration.  The research has made use of the NASA/IPAC Infrared
Science Archive, which is operated by the Jet Propulsion Laboratory,
California Institute of Technology, under contract with the National
Aeronautics and Space Administration.

\bsp

\label{lastpage}


\begin{thebibliography}{99}

\bibitem[\protect\citeauthoryear{Bachiller et al.}{1990}]{bac90}
Bachiller R., Menten K.M., G\'omez-Gonz\'alez J., Barcia A., 1990, A\&A, 240, 116

\bibitem[\protect\citeauthoryear{Batrla et al.}{1987}]{bat87}
Batrla W., Matthews H.E., Menten K.M., Walmsley C.M., 1987, Nature, 326, 49

\bibitem[\protect\citeauthoryear{Beuther et al.}{2005}]{beu05}
Beuther H., Thorwirth S., Zhang Q., Hunter T.R., Megeath S.T., Walsh A.J., 
Menten K.M., 2005, ApJ, 627, 834

\bibitem[\protect\citeauthoryear{Beuther et al.}{2008}]{beu08}
Beuther H., Walsh A.J., Thorwirth S., Zhang Q., Hunter T.R., Megeath S.T., 
Menten K.M., 2008, A\&A, 481, 169

\bibitem[\protect\citeauthoryear{Breen et al.}{2010}]{bre10}
Breen S.L., Ellingsen S.P., Caswell J.L., Lewis B.E., 2010, MNRAS, 401, 2219 (arXiv:0910.1223)

\bibitem[\protect\citeauthoryear{Brogan et al.}{2009}]{bro09}
Brogan C.L., Hunter T.R., Cyganowski C.J., Indebetouw R., Beuther H., Menten K.M., 
Thorwirth S., 2009, ApJ, 707, 1

\bibitem[\protect\citeauthoryear{Brooks et al.}{2003}]{bro03}
Brooks K.J., Garay G., Mardones D., Bronfman L., 2003, ApJ, 594, 131

\bibitem[\protect\citeauthoryear{Caswell \& Haynes}{1987}]{cas87}
Caswell J.L., Haynes R.F., 1987, Aust. J. Phys., 40, 215

\bibitem[\protect\citeauthoryear{Caswell et al.}{1995}]{cas95}
Caswell J.L., Vaile R.A., Ellingsen S.P., Whiteoak J.B., Norris R.P., 1995, MNRAS, 272, 96

\bibitem[\protect\citeauthoryear{Caswell}{1997}]{cas97}
Caswell J.L., 1997, MNRAS, 289, 203

\bibitem[\protect\citeauthoryear{Caswell}{1998}]{cas98}
Caswell J.L., 1998, MNRAS, 297, 215

\bibitem[\protect\citeauthoryear{Caswell}{2009}]{cas09}
Caswell J.L., 2009, PASA, 26, 454 (arXiv:0907.5255)

\bibitem[\protect\citeauthoryear{Caswell et al.}{2010}]{cas10}
Caswell J.L., Fuller G.A., Green J.A., Avison A., Breen S.L., Brooks K.J., 
Burton M.G., Chrysostomou A., Cox J., Diamond P.J., Ellingsen S.P.,
Gray M.D., Hoare M.G., Masheder M.R.W., McClure-Griffiths N.M., 
Pestalozzi M.R., Phillips C.J., Quinn L., Thompson M.A., Voronkov M.A., 
Walsh A.J., Ward-Thompson D., Wong-McSweeney D., Yates J.A., Cohen R.J.,
2010, MNRAS, in press (arXiv:1002.2475)

\bibitem[\protect\citeauthoryear{Chen, Ellingsen \& Shen}{Chen et al.}{2009}]{che09}
Chen X., Ellingsen S.P., Shen Z.-Q., 2009, MNRAS, 396, 1603

\bibitem[\protect\citeauthoryear{Codella \& Moscadelli}{2000}]{cod00}
Codella C., Moscadelli L., 2000, A\&A, 362, 723

\bibitem[\protect\citeauthoryear{Cragg et al.}{1992}]{cra92}
Cragg D.M., Johns K.P., Godfrey P.D., Brown R.D., 1992, MNRAS, 259, 203

\bibitem[\protect\citeauthoryear{Cragg et al.}{2004}]{cra04}
Cragg D.M., Sobolev A.M., Caswell J.L., Ellingsen S.P., Godfrey P.D., 
2004, MNRAS, 351, 1327

\bibitem[\protect\citeauthoryear{Cyganowski et al.}{2008}]{cyg08}
Cyganowski C.J., Whitney B.A., Holden E., Braden E., Brogan C.L., Churchwell E.,
Indebetouw R., Watson D.F., Babler B.L., Benjamin R., Gomez M., Meade M.R., 
Povich M.S., Robitaille T.P., Watson C., 2008, AJ, 136, 2391

\bibitem[\protect\citeauthoryear{Cyganowski et al.}{2009}]{cyg09}
Cyganowski C.J., Brogan C.L., Hunter T.R., Churchwell E., ApJ, 2009, 702, 1615

\bibitem[\protect\citeauthoryear{Davis et al.}{2004}]{dav04}
Davis C.J., Varricatt W.P., Todd S.P., Ramsay Howat S.K., 2004, A\&A, 425, 981

\bibitem[\protect\citeauthoryear{De Buizer}{2003}]{deb03}
De Buizer J.M., 2003, MNRAS, 341, 277

\bibitem[\protect\citeauthoryear{De Buizer et al.}{2009}]{deb09}
De Buizer J.M., Redman R.O., Longmore S.N., Caswell J., Feldman P.A.,
2009, A\&A, 493, 127 

\bibitem[\protect\citeauthoryear{Ellingsen et al.}{2004}]{ell04}
Ellingsen S.P., Cragg D.M., Lovell J.E.J., Sobolev A.M., Ramsdale P.D., Godfrey P.D.,
2004, MNRAS, 354, 401

\bibitem[\protect\citeauthoryear{Ellingsen}{2005}]{ell05}
Ellingsen S.P., 2005, MNRAS, 359, 1498

\bibitem[\protect\citeauthoryear{Ellingsen}{2006}]{ell06}
Ellingsen S.P., 2006, ApJ, 638, 241

\bibitem[\protect\citeauthoryear{Ellingsen et al.}{2007}]{ell07}
Ellingsen S.P., Voronkov M.A., Cragg D.M., Sobolev A.M., Breen S.L., Godfrey P.D.,
2007,  proceedings of IAU Symposium 242, Astrophysical Masers and their Environments
(eds. J. M. Chapman, W. A. Baan), 213 (arXiv:0705.2906)

\bibitem[\protect\citeauthoryear{Forster \& Caswell}{1989}]{for89}
Forster J.R., Caswell J.L., 1989, A\&A, 213, 339

\bibitem[\protect\citeauthoryear{Furuya et al.}{2005}]{fur05}
Furuya R.S., Cesaroni R., Takahashi S., Momose M., 
Testi L., Shinnaga H., Codella C., 2005, ApJ, 624, 827

\bibitem[\protect\citeauthoryear{Garay, Reid \& Moran}{Garay et al.}{1985}]{gar85}
Garay G., Reid M.J., Moran J.M., 1985, ApJ, 289, 681

\bibitem[\protect\citeauthoryear{Garay et al.}{1998}]{gar98}
Garay G., Moran J.M., Rodr\'\i guez L.F., Reid M.J., 1998, ApJ, 492, 635

\bibitem[\protect\citeauthoryear{Garay et al.}{2003}]{gar03}
Garay G., Brooks K.J., Mardones D., Norris R.P., 2003, ApJ, 587, 739

\bibitem[\protect\citeauthoryear{Gibb \& Davis}{1998}]{gib98}
Gibb A.G., Davis C.J., 1998, MNRAS, 298, 644

\bibitem[\protect\citeauthoryear{Goedhart, Gaylard \& van der Walt}
{Goedhart et al.}{2004}]{goe04} Goedhart S., Gaylard M.J., van der Walt D.J.,
2004, MNRAS, 355, 553

\bibitem[\protect\citeauthoryear{Goldsmith \& Mao}{1983}]{gol83}
Goldsmith P.F., Mao X.J., 1983, ApJ, 265, 791

\bibitem[\protect\citeauthoryear{Goss, Matthews \& Winnberg}{Goss et al.}{1978}]{gos78}
Goss W.M., Matthews H.E., Winnberg A., 1978, A\&A, 65, 307

\bibitem[\protect\citeauthoryear{Green et al.}{2009}]{gre09}
Green J.A., Caswell J.L., Fuller G.A., Breen S.L., Brooks K., Burton M.G., Chrysostomou A.,
Cox J., Diamond P.J., Ellingsen S.P., Gray M.D., Hoare M.G., Masheder M.R.W., 
McClure-Griffiths N., Pestalozzi M., Phillips C., Quinn L., Thomson M.A., Voronkov M.A.,
Walsh A., Ward-Thompson D., Wong-McSweeney, Yates J.A., Cohen R.J., 2009,
MNRAS, 392, 783

\bibitem[\protect\citeauthoryear{Haschick, Menten \& Baan}{Haschick et al.}{1990}]{has90}
Haschick A.D., Menten K.M., Baan W.A., 1990, ApJ, 354, 556

\bibitem[\protect\citeauthoryear{Henning et al.}{2000}]{hen00}
Henning Th., Lapinov A., Schreyer K., Stecklum B., Zinchenko~I.,
2000, A\&A, 364, 613 

\bibitem[\protect\citeauthoryear{Hill et al.}{2005}]{hil05}
Hill T., Burton M.G., Minier V., Thompson M.A., Walsh A.J., Hunt-Cunninham M., Garay G.,
2005, MNRAS, 363, 405

\bibitem[\protect\citeauthoryear{Ho, Klein \& Haschick}{Ho et al.}{1986}]{ho86}
Ho P.T.P., Klein R.I., Haschick A.D.,  1986, ApJ,  305, 714

\bibitem[\protect\citeauthoryear{Keto \& Ho}{1989}]{ket89}
Keto E.R., Ho P.T.P., 1989, ApJ, 347, 349

\bibitem[\protect\citeauthoryear{Kirsanova, Wiebe \& Sobolev}{Kirsanova et al.}{2009}]{kir09}
Kirsanova M.S., Wiebe D.S., Sobolev A.M., 2009, Astron. Rep., 2009, 53, 611

\bibitem[\protect\citeauthoryear{Kogan \& Slysh}{1998}]{kog98}
Kogan L., Slysh V., 1998, ApJ, 497, 800

\bibitem[\protect\citeauthoryear{Kurtz, Hofner \& \'Alvarez}
{Kurtz et al.}{2004}]{kur04}
Kurtz S., Hofner P., \'Alvarez C.V., 2004, ApJSS, 155, 149

\bibitem[\protect\citeauthoryear{Lees}{1973}]{lee73}
Lees R.M., 1973, ApJ, 184, 763

\bibitem[\protect\citeauthoryear{Liechti \& Wilson}{1996}]{lie96}
Liechti S., Wilson T.L., 1996, A\&A, 314, 615

\bibitem[\protect\citeauthoryear{Mehringer}{1994}]{meh94}
Mehringer D.M., 1994, ApJS, 91, 713

\bibitem[\protect\citeauthoryear{Mehringer \& Menten}{1996}]{meh96}
Mehringer D.M., Menten K.M., 1996, ApJ, 474, 346

\bibitem[\protect\citeauthoryear{Menten et al.}{1986}]{men86}
Menten K.M., Walmsley C.M., Henkel C., Wilson T.L., 1986, A\&A, 157, 318

\bibitem[\protect\citeauthoryear{Menten \& Batrla}{1989}]{men89}
Menten K.M., Batrla W., 1989, ApJ, 341, 839

\bibitem[\protect\citeauthoryear{Menten}{1991}]{men91}
Menten K.M., 1991, ApJ, 380, L75

\bibitem[\protect\citeauthoryear{M\"uller, Menten \& M\"ader}
{M\"uller et al.}{2004}]{mul04}
M\"uller H.S.P., Menten K.M., M\"ader H., 2004, A\&A, 428, 1019

\bibitem[\protect\citeauthoryear{Mu\~nos et al.}{2007}]{mun07}
Mu\~nos D.J., Mardones D., Garay G., Rebolledo D., Brooks K., Bontemps S.,
2007, ApJ, 668, 906

\bibitem[\protect\citeauthoryear{Norris et al.}{1993}]{nor93}
Norris R.P., Whiteoak J.B., Caswell J.L., Wieringa M.H., Gough R.G.,
1993, ApJ, 412, 222

\bibitem[\protect\citeauthoryear{Phillips et al.}{1998}]{phi98}
Phillips C.J., Norris R.P., Ellingsen S.P., McCulloch P.M., 1998, MNRAS, 300, 1131

\bibitem[\protect\citeauthoryear{Plambeck \& Menten}{1990}]{pla90}
Plambeck R.L., Menten K.M., 1990, ApJ, 364, 555

\bibitem[\protect\citeauthoryear{Pratap et al.}{2008}]{pra08}
Pratap P., Shute P.A., Keane T.C., Battersby C., Sterling S., 2008, ApJ, 135, 1718

\bibitem[\protect\citeauthoryear{Rodr\'\i guez et al.}{2005}]{rod05}
Rodr\'\i guez L.F., Garay G., Brooks K.J., Mardones D., 2005, ApJ, 626, 953

\bibitem[\protect\citeauthoryear{Salii, Sobolev \& Kalinina}
{Salii et al.}{2002}]{sal02}
Salii S.V., Sobolev A.M., Kalinina N.D., 2002, Astron. Rep., 46, 955

\bibitem[\protect\citeauthoryear{Slysh, Kalenskii \& Val'tts}{Slysh et al.}{1993}]{sly93}
Slysh V.I., Kalenskii S.V., Val'tts I.E., 1993, ApJ, 413, 133

\bibitem[\protect\citeauthoryear{Slysh et al.}{1994}]{sly94}
Slysh V.I., Kalenskii S.V., Val'tts I.E., Otrupcek R., 1994, MNRAS, 268, 464

\bibitem[\protect\citeauthoryear{Slysh et al.}{1999}]{sly99}
Slysh V.I., Val'tts I.E., Kalenskii S.V., Voronkov M.A., Palagi F.,
Tofani G., Catarzi M., 1999, A\&ASS, 134, 115

\bibitem[\protect\citeauthoryear{Sobolev \& Strelnitskii}{1983}]{sob83}
Sobolev A.M., Strelnitskii V.S., 1983, Soviet  Astron. Letters, 9, 12

\bibitem[\protect\citeauthoryear{Sobolev}{1992}]{sob92}
Sobolev A.M., 1992, Soviet Astron., 36, 590

\bibitem[\protect\citeauthoryear{Sobolev \& Deguchi}{1994}]{sob94}
Sobolev A.M., Deguchi S., 1994, A\&A, 291, 569

\bibitem[\protect\citeauthoryear{Sobolev et al.}{2005}]{sob05}
Sobolev A.M., Ostrovskii A.B., Kirsanova M.S., Shelemei O.V.,
Voronkov M.A., Malyshev A.V., 2005, proceedings of IAU Symposium 227
(eds. E.Churchwell, P.Conti and M.Felli), 174 (astro-ph/0601260)

\bibitem[\protect\citeauthoryear{Sutton et al.}{2001}]{sut01}
Sutton E.C., Sobolev A.M., Ellingsen S.P., Cragg D.M., Mehringer D.M.,
Ostrovskii A.B., Godfrey P.D., 2001, ApJ, 554, 173

\bibitem[\protect\citeauthoryear{Sutton et al.}{2004}]{sut04}
Sutton E.C., Sobolev A.M., Salii S.V., Malyshev A.V., Ostrovskii A.B., Zinchenko I.I.,
2004, ApJ, 609, 231

\bibitem[\protect\citeauthoryear{Szymczak, Pillai \& Menten}{Szymczak et al.}{2005}]{szy05}
Szymczak M., Pillai T., Menten K.M., 2005, A\&A, 434, 613


\bibitem[\protect\citeauthoryear{Val'tts et al.}{2000}]{val00}
Val'tts I.E., Ellingsen S.P., Slysh V.I., Kalenskii S.V.,
Otrupcek R., Larionov G.M., 2000, MNRAS, 317, 315

\bibitem[\protect\citeauthoryear{Voronkov}{1999}]{vor99}
Voronkov M.A., 1999, Astron. Lett., 25, 149 (astro-ph/0008476)

\bibitem[\protect\citeauthoryear{Voronkov et al.}{2005a}]{vor05a}
Voronkov M.A., Sobolev A.M., Ellingsen S.P., Ostrovskii A.B., Alakoz A.V., 2005a,
Ap\&SS, 295, 217

\bibitem[\protect\citeauthoryear{Voronkov et al.}{2005b}]{vor05b}
Voronkov M.A., Sobolev A.M., Ellingsen S.P., Ostrovskii A.B.,
2005b, MNRAS, 362, 995

\bibitem[\protect\citeauthoryear{Voronkov et al.}{2006}]{vor06}
Voronkov M.A., Brooks K.J., Sobolev A.M., Ellingsen S.P., Ostrovskii A.B., Caswell J.L., 2006, 
MNRAS, 373, 411

\bibitem[\protect\citeauthoryear{Voronkov et al.}{2007}]{vor07}
Voronkov M.A., Brooks K.J., Sobolev A.M., Ellingsen S.P., Ostrovskii A.B., Caswell J.L., 2007,
proceedings of IAU Symposium 242, Astrophysical Masers and their Environments
(eds. J. M. Chapman, W. A. Baan), 182 (arXiv:0705.0355)

\bibitem[\protect\citeauthoryear{Walsh et al.}{1998}]{wal98}
Walsh A.J., Burton M.G., Hyland A.R., Robinson G., 1998, MNRAS, 301, 640

\bibitem[\protect\citeauthoryear{Walsh et al.}{2003}]{wal03}
Walsh A.J., Macdonald G.H., Alvey N.D.S., Burton M.G., Lee J.-K., 2003, A\&A, 410, 597

\bibitem[\protect\citeauthoryear{Walsh et al.}{2007}]{wal07}
Walsh A.J., Chapman J.F., Burton M.G., Wardle M., Millar T.J., 2007, MNRAS, 380, 1703

\bibitem[\protect\citeauthoryear{Wilson et al.}{1985}]{wil85}
Wilson T.L., Walmsley C.M., Menten K.M., Hermsen W., 1985, A\&A, 147, L19

\bibitem[\protect\citeauthoryear{Wu et al.}{2009}]{wu09}
Wu Y., Qin S.-L., Guan X., Xue R., Ren Z., Liu T., Huang M., Chen S., 2009, ApJ, 697, L116

\end{thebibliography}
\end{document}